\documentclass[11pt, letterpaper]{article}
\usepackage{amsfonts,amssymb,amsmath}            
\usepackage{graphics,graphicx,epsfig}            
\usepackage{amsthm}                              

\newlength{\protwidthl}
\newlength{\protwidthm}
\newlength{\protwidthr}
\newlength{\protarrowraise}
\newlength{\protarrowlength}
\newcommand{\protdefscale}{.8}

\newcommand{\genarrow}[3]%
{ \setlength{\unitlength}{#1}%
  \raisebox{0.5ex}{\begin{picture}(0,0)\thinlines #3\end{picture}}%
  \raisebox{1.4ex}{\makebox[#1]{#2}}%
}

\newcommand{\vb}[1]{\rule[#1ex]{0pt}{0pt}}

\newcommand{\protright}[3]%
{ \parbox[b]{\protwidthl}{\raggedright\vb{2.5}\mbox{}#1}
  & \genarrow{\protarrowlength}{#2}%
    {\put(0,0){\vector(1,0){1}}}
  & \parbox[t]{\protwidthr}{\raggedright\mbox{}#3\vb{-2}}\\}
\newcommand{\protleft}[3]%
{ \parbox[t]{\protwidthl}{\raggedright\mbox{}#1\vb{-2}}
  & \genarrow{\protarrowlength}{#2}%
    {\put(1,0){\vector(-1,0){1}}}
  & \parbox[b]{\protwidthr}{\raggedright\vb{2.5}\mbox{}#3}\\}
\newcommand{\protno}[2]%
{ \parbox[t]{\protwidthl}{\raggedright\vb{2.5}\mbox{}#1\vb{-2}}
  && \parbox[t]{\protwidthr}{\raggedright\vb{2.5}\mbox{}#2\vb{-2.5}}\\}
\newcommand{\protcenter}[1]{\hfill #1\hfill\hfill}

\newenvironment{protocol}[3][\protdefscale]%
{ \setlength{\protwidthl}{#1\linewidth}%
  \setlength{\protwidthm}{.25\protwidthl}%
  \setlength{\protwidthr}{.375\protwidthl}%
  \setlength{\protwidthl}{\protwidthr}%
  \setlength{\protarrowraise}{1.5ex}%
  \setlength{\protarrowlength}{\protwidthm}%
  \addtolength{\protarrowlength}{-.1\protarrowlength}%
  \begin{center}
    \begin{tabular}{p{\protwidthl}cp{\protwidthr}}
      \protcenter{\bf #2} && \protcenter{\bf #3}\vspace{.5ex}\\
}%
{   \end{tabular}
  \end{center}
}

{ \setlength{\protwidthl}{#1}%
  \setlength{\protwidthm}{#2}%
  \setlength{\protwidthr}{#3}%
  \setlength{\protarrowraise}{1.5ex}%
  \setlength{\protarrowlength}{\protwidthm}%
  \addtolength{\protarrowlength}{-.1\protarrowlength}%
  \begin{center}
    \begin{tabular}{p{\protwidthl}cp{\protwidthr}}
      \protcenter{\bf #4} && \protcenter{\bf #5}\vspace{.5ex}\\
}%
{   \end{tabular}
  \end{center}
}

\begin{document}



\def\openone{\leavevmode\hbox{\small1\normalsize\kern-.33em1}}
\newcommand{\proj}[1]{\ket{#1}\bra{#1}}
\newcommand{\braket}[2]{\left\langle\, #1\,|\,#2\,\right\rangle}
\newcommand{\outprod}[2]{\ket{#1}\bra{#2}}
\newcommand{\half}{\mbox{$\textstyle \frac{1}{2}$}}
\newcommand{\h}[1]{\mathcal{H}_{#1} }
\newcommand{\be}{\begin{equation}}
\newcommand{\ee}{\end{equation}}
\newcommand{\bea}{\begin{eqnarray}}
\newcommand{\eea}{\end{eqnarray}}
\newcommand{\rem}[1]{\marginpar{\begin{flushleft} \scriptsize
#1\end{flushleft}}}
\bibliographystyle{unsrt}

\def\opone{\leavevmode\hbox{\small1\kern-3.8pt\normalsize1}}
\newcommand{\re}{\mbox{Re}}





\theoremstyle{plain}
\newtheorem{theorem}{Theorem}[section]
\newtheorem{lemma}[theorem]{Lemma}
\newtheorem{corollary}[theorem]{Corollary}
\theoremstyle{definition}
\newtheorem{definition}[theorem]{Definition}

\newcommand*{\ab}{\bar{a}}
\newcommand*{\Ab}{\bar{A}}
\newcommand*{\Fb}{\bar{F}}
\newcommand*{\Ph}{\hat{P}}
\newcommand*{\Pp}{Q}
\newcommand*{\Qh}{\hat{Q}}
\newcommand*{\xb}{\bar{x}}
\newcommand*{\zb}{\bar{z}}
\newcommand*{\zp}{z'}
\newcommand*{\Rp}{R'}
\newcommand*{\zh}{\hat{z}}
\newcommand*{\Zh}{\hat{Z}}
\newcommand*{\Zp}{Z'}
\newcommand*{\deltab}{\bar{\delta}}
\newcommand*{\eps}{\varepsilon}
\newcommand*{\epsp}{\varepsilon'}
\newcommand*{\epspp}{\varepsilon''}
\newcommand*{\epsb}{\bar{\varepsilon}}
\newcommand*{\rhoh}{\hat{\rho}}
\newcommand*{\mub}{\bar{\mu}}

\newcommand*{\bbC}{\mathbb{C}}
\newcommand*{\bbN}{\mathbb{N}}
\newcommand*{\bbR}{\mathbb{R}}

\newcommand*{\bp}{\mathbf{p}}
\newcommand*{\bw}{\mathbf{w}}
\newcommand*{\bW}{\mathbf{W}}
\newcommand*{\bx}{\mathbf{x}}
\newcommand*{\bX}{\mathbf{X}}
\newcommand*{\bXb}{\mathbf{\bar{X}}}
\newcommand*{\bXh}{\mathbf{\hat{X}}}
\newcommand*{\by}{\mathbf{y}}
\newcommand*{\bY}{\mathbf{Y}}
\newcommand*{\bYh}{\mathbf{\hat{Y}}}
\newcommand*{\bz}{\mathbf{z}}
\newcommand*{\bzb}{\mathbf{\bar{z}}}
\newcommand*{\bzp}{\mathbf{z'}}
\newcommand*{\bZ}{\mathbf{Z}}
\newcommand*{\bZb}{\mathbf{\bar{Z}}}
\newcommand*{\bZp}{\mathbf{Z'}}
\newcommand*{\bZh}{\mathbf{\hat{Z}}}
\newcommand*{\bZt}{\mathbf{\tilde{Z}}}

\newcommand*{\cA}{\mathcal{A}}
\newcommand*{\cB}{\mathcal{B}}
\newcommand*{\cC}{\mathcal{C}}
\newcommand*{\cE}{\mathcal{E}}
\newcommand*{\cF}{\mathcal{F}}
\newcommand*{\cFb}{\bar{\mathcal{F}}}
\newcommand*{\cG}{\mathcal{G}}
\newcommand*{\cH}{\mathcal{H}}
\newcommand*{\cI}{\mathcal{I}}
\newcommand*{\cJ}{\mathcal{J}}
\newcommand*{\cM}{\mathcal{M}}
\newcommand*{\cP}{\mathcal{P}}
\newcommand*{\cQ}{\mathcal{Q}}
\newcommand*{\cR}{\mathcal{R}}
\newcommand*{\cS}{\mathcal{S}}
\newcommand*{\cT}{\mathcal{T}}
\newcommand*{\cU}{\mathcal{U}}
\newcommand*{\cV}{\mathcal{V}}
\newcommand*{\cW}{\mathcal{W}}
\newcommand*{\cX}{\mathcal{X}}
\newcommand*{\cY}{\mathcal{Y}}
\newcommand*{\cZ}{\mathcal{Z}}
\newcommand*{\cZb}{\mathcal{\bar{Z}}}
\newcommand*{\cZs}{\mathcal{Z'}}

\newcommand*{\setA}{a}
\newcommand*{\setAb}{\bar{a}}
\newcommand*{\setB}{b}
\newcommand*{\setI}{\mathcal{I}}

\newcommand*{\fmax}[1]{#1^{\max}}  
\newcommand*{\fmin}[1]{#1^{\min}}  

\newcommand*{\Prob}{\mathrm{Prob}}
\newcommand*{\POVM}{\mathrm{POVM}}

\newcommand*{\freq}[1]{Q_{#1}} 

\newcommand*{\qH}{S} 

\newcommand*{\event}{\mathcal{E}}
\newcommand*{\eventb}{\bar{\mathcal{E}}}

\newcommand*{\ket}[1]{| #1 \rangle}
\newcommand*{\bra}[1]{\langle #1 |}
\newcommand*{\spr}[2]{\langle #1 | #2 \rangle}
\newcommand*{\tr}{\mathrm{tr}}
\newcommand*{\id}{\mathrm{id}}
\newcommand*{\rank}{\mathrm{rank}}

\newcommand*{\ev}{\boldsymbol{\lambda}}

\newcommand*{\sbin}{\{0,1\}} 
\newcommand*{\dist}{\delta} 
\newcommand*{\Pbin}[1]{P^{\mathrm{bin}}_{#1}} 
\newcommand*{\meas}{\gamma} 
\newcommand*{\ZVmeas}{\Gamma} 

\newcommand*{\tu}{two-universal}
\newcommand*{\tuty}{\tu ity}

\title{A Generic Security Proof for Quantum Key Distribution}

\author{Matthias Christandl
\thanks{Centre for Quantum Computation,
             Department of Applied Mathematics and Theoretical Physics,
             University of Cambridge,
             Wilberforce Road,
             Cambridge CB3 0WA, United Kingdom}
\and Renato Renner \thanks{Computer Science Department,
             ETH Z\"urich,
             CH-8092 Z\"urich,
             Switzerland}
\and Artur Ekert {\addtocounter{footnote}{-2}\footnotemark}
{\addtocounter{footnote}{1}}
\thanks{Department of Physics,
             National University of Singapore,
             Singapore 117\,542, Singapore}}

\date{March 4, 2004}
%
%
%
%
\maketitle
%
%

\begin{abstract}
Quantum key distribution allows two parties, traditionally known as Alice and
Bob, to establish a secure random cryptographic key if, firstly, they have
access to a quantum communication channel, and secondly, they can exchange
classical public messages which can be monitored but not altered by an
eavesdropper, Eve. Quantum key distribution provides perfect security
because, unlike its classical counterpart, it relies on the laws of physics
rather than on ensuring that successful eavesdropping would require excessive
computational effort. However, security proofs of quantum key distribution
are not trivial and are usually restricted in their applicability to specific
protocols. In contrast, we present a general and conceptually simple proof
which can be applied to a number of different protocols. It relies on the
fact that a cryptographic procedure called privacy amplification is equally
secure when an adversary's memory for data storage is quantum rather than
classical~\cite{KMR03}.
\end{abstract}


\section{Introduction}


The potential power of quantum phenomena to protect information was first
adumbrated by Wiesner who, in the early 1970's, introduced the concept of
quantum conjugate coding~\cite{Wie70}. He showed how to store or transmit two
messages by encoding them in two conjugate observables, such as linear and
circular polarization of light, so that either but not both of may be
received and decoded. He illustrated his idea with a design of unforgeable
bank notes. Building upon this work, Bennett and Brassard proposed a quantum
key distribution scheme, known as BB84 or the four state protocol, in which
Alice repeatedly sends to Bob one of four prescribed states of a qubit, and
Bob measures them in one of two conjugate bases~\cite{BB84}. Independently
and initially unaware of the earlier work, Ekert developed a different
approach to quantum cryptography based on quantum entanglement. He proposed a
key distribution protocol, known as E91, in which entangled pairs of qubits
are distributed to Alice and Bob, who then extract key bits by measuring
their qubits in prescribed bases~\cite{E91}. A particularly nice feature of
E91, for the purpose of security analysis, is that Eve herself is allowed to
prepare and deliver all the qubit pairs that Alice and Bob will subsequently
use to generate the key.

Many variations on quantum key distribution have been subsequently proposed
and we will mention some of them later on. They can be roughly divided into
``prepare and measure" protocols, such as BB84 and B92~\cite{Ben92}, and
``entanglement based" protocols, such as E91. Many interesting techniques for
manipulating quantum entanglement have been discovered in the last few years.
Thus it is often convenient to cast some of the ``prepare and measure"
protocols in terms of the ``entanglement based" ones.

\subsection{Security Proofs}
All good quantum key distribution protocols must be operable in the presence
of noise that may or may not result from eavesdropping. The protocols must
specify for which values of measurable parameters Alice and Bob can establish
a secret key and provide a physically implementable procedure which generates
such a key. The design of the procedure must take into account that an
eavesdropper may have access to unlimited quantum computing power. On Alice
and Bob's side, the procedure should rely on simple and easily implementable
operations. For example, good protocols should not assume that Alice and Bob
have quantum computers, or any sophisticated quantum technology, apart from
the ability to transmit over a quantum channel.

The search for operational security criteria led to early studies of quantum
eavesdropping~\cite{EH94,Yao95} and finally to the first proof of the
security of key distribution~\cite{DEJMPS96}. The original proof showed that
the E91 and all entanglement based key distributions are indeed secure and
noise-tolerant against an adversary with unlimited computing power as long as
Alice and Bob can implement quantum privacy amplification. Quantum privacy
amplification allows one to establish a secure key over any distance, e.g.
using entanglement swapping~\cite{ZZHE93} in a chain of quantum
repeaters~\cite{DBCZ99,AB02}. However, this procedure, which distills pure
entangled states from corrupted mixed states of two qubits, requires a small
scale quantum computation. Subsequent proofs by Inamori~\cite{Ina00} and
Ben-Or~\cite{BO02} showed that Alice and Bob can also distill a secret key
from partially entangled particles using only classical error correction and
classical privacy amplification~\cite{BBR88,BBCM95}.

Quantum privacy amplification was also used by Lo and Chau to prove the
security of the BB84 protocol over an arbitrary distance~\cite{LC99}. A
concurrent and independent proof by Mayers showed that the protocol can be
secure without Alice and Bob having to rely on the use of quantum
computers~\cite{May98}. The same conclusion, but using different techniques,
was subsequently reached by Biham {\it et al.}~\cite{BBBMR00}. Although the
two proofs did not require quantum privacy amplification they were rather
complex. A nice fusion of quantum privacy amplification and error correction
was proposed by Shor and Preskill who formulated a relatively simple proof of
the security of the BB84 protocol based on virtual quantum error
correction~\cite{SP00}. They showed that a protocol which employs quantum
error-correcting codes to prevent Eve from becoming entangled with qubits
that are used to generate the key reduces to the BB84 augmented by classical
error correction and classical privacy amplification. This proof has been
further extended by Gottesman and Lo~\cite{GL03} to cover the case of two-way
public communication in BB84 which allows a higher bit error rate, and by
Tamaki {\it et al.}\cite{TKI03} to prove the security of the B92 protocol.
More recently another simple proof of the security of BB84, which employs
results from quantum communication complexity, has been provided by
Ben-Or~\cite{BO02}.

\subsection{Do we need another Security Proof?}

Most popular quantum key distribution schemes have been analyzed in
terms of their security criteria and there is a pretty good
understanding of the limitations of the techniques involved e.g. those
due to imperfect sources or detectors. The schemes vary but every
single one of them must involve either quantum or classical privacy
amplification as an inherent part of the secure key distillation
protocol.

Classical privacy amplification, originally proposed by Bennett,
Brassard and Robert, was restricted to the case in which Eve acquires
\emph{classical, deterministic} information about the raw
key~\cite{BBR88}. The applicability of the method was then extended by
Bennett, Brassard, Cr\'epeau and Maurer to cover scenarios where Eve's
information is \emph{classical} and
\emph{probabilistic}~\cite{BBCM95}. We use the recent result by
K\"onig, Maurer and Renner on the power of quantum memory~\cite{KMR03}
in a quantum cryptographic context. It can be viewed as a further
generalization of classical privacy amplification to cases in which
Eve's information about the key is \emph{quantum}. Of course, the
privacy amplification is useless unless we can derive an upper bound
on the amount of quantum information available to Eve. We show how to
do this for common quantum key distribution protocols.  Taken together
these results give a very general and powerful technique for assessing
security of a wide class of quantum key distribution protocols.

\subsection{Scenario}

In our scenario Eve has a technological advantage over Alice and Bob. She can
distribute qubits to Alice and Bob, she can entangle the qubits with an
ancilla that she controls, she can have access to unlimited quantum
computational power, and she can monitor all the public communication between
Alice and Bob in which they reveal their measurement choices and exchange
further information in order to correct errors in their shared key and to
amplify its privacy. In contrast Alice and Bob can only perform measurements
on individual qubits and communicate classically over a public channel. We
will assess the security in the case of a noisy quantum channel without
losses.

Alice and Bob go through prescribed stages of quantum key distribution
and at some point they end up with perfectly correlated binary strings
about which Eve has some information, namely all information
communicated in public together with all information contained in her
ancilla. The ancilla is a quantum entity which Eve may measure at the
very end of the key distribution protocol. Hence its information
content has to be expressed in qubits rather than bits. Classical
privacy amplification allows Alice and Bob to increase the privacy of
the shared string as long as they can estimate the amount of classical
information that leaked to Eve~\cite{BBR88}. For any shared string of
$n$ bits upon which Eve has some $r$ bits of information the procedure
outputs a binary string of length $s$ shorter than $n-r$ and such that
Eve has virtually no information about the new string. The snag is
that Eve, who can delay her measurement of the ancilla, has $r$ qubits
rather than $r$ bits of information about the $n$--bit string.
However, in this particular context, it does not matter, as shown
in~\cite{KMR03}. We show how Alice and Bob are able to estimate the
quantum information content of $r$ qubits in Eve's ancilla in a
generic quantum key distribution protocol.

\section{Outline of the Main Result}

It is convenient for the purpose of this outline to start with a generic
scenario in which Eve distributes quantum particles to Alice and Bob. Without
any loss of generality we assume that Eve starts with a tripartite pure state
describing a batch of particles delivered to Alice, a batch of particles
delivered to Bob, and an ancilla which is retained by Eve.

When Alice and Bob receive their respective particles they perform
measurements following a quantum key distribution protocol, which they agreed
to in advance. For example, they may measure every single particle choosing
randomly from a prescribed set of different measurements. They also
communicate in public and agree which outcomes of the measurements are to be
discarded and which will be used for the key generation.

At this point Alice and Bob have partially correlated $n$ bit strings
labeled, respectively, as $X$ and $Y$. Eve knows the protocol and holds an
ancilla which was entangled with the qubits prior to Alice's and Bob's
measurements. After the measurements the ancilla is in a quantum state which,
in general, depends on $X$ and $Y$ and is described by some density operator
$\rho^E$. The initial public communication must allow Alice and Bob to
estimate the degree of the correlation between $X$ and $Y$ and to derive an
upper bound on the quantum information content of the ancilla in state
$\rho^E$. This is not trivial as we do not assume that the pairs of qubits
are independent and identically distributed (i.i.d); they can be entangled
between themselves and the ancilla in an arbitrary way.

We solve the problem in its full generality. However, in this section we
present a rough outline based on the i.i.d case. This, we hope, will serve as
a gentle introduction to the more technical sections that follow.

Let Alice and Bob be given $n$ realizations of i.i.d random variables
$X$ and $Y$ respectively. Let the degree of correlations be quantified
by the mutual information $I(X;Y)$ and let the quantum information
content of the ancilla be no more than $r$ qubits. The strings of
Alice and Bob can be made identical with high probability by a
procedure called information reconciliation. Alice has to communicate
in public approximately $n H(X|Y)$ (the conditional entropy of $X$
given $Y$) bits about her string so that Bob, who holds $n$
realizations of $Y$, can guess Alice's string correctly.

Thus, after the information reconciliation, Eve's information about Alice's
string consists of $n H(X|Y)$ classical bits and $r$ qubits. Without any loss
of generality we can assume that Eve's information is contained in $n
H(X|Y)+r$ qubits. Eve can wait and perform her measurement on the ancilla
whenever she sees fit. However, no matter which observable she measures after
the classical privacy amplification she is not better off than she would be
if she had $n H(X|Y)+r$ classical bits of information about $X$ prior to the
privacy amplification. This follows from the recent work by K\"onig, Maurer,
and Renner on the power of quantum memory~\cite{KMR03}. We will elaborate on
this in more detail in section \ref{sec:preliminaries} and section
\ref{sec:main}. Thus the length of the secret key after the privacy
amplification is $n H(X)-n H(X|Y)-r=n I(X;Y)-r$, i.e. the key can be
established when $n I(X;Y)> r$.

In the main part of the paper we will show how the estimation of $r$ works in
general. In order to illustrate the idea behind this estimation, let us
consider the particular case of independent and identically distributed pairs
of quantum states. Each pair that Eve delivers to Alice and Bob comes from a
tripartite pure state $\ket{\Psi}$ such that $\rho = \tr_E\proj{\Psi}$ is the
density operator of each pair of quantum states and $\rho^E =
\tr_{AB}\proj{\Psi}$ is the density operator of a part of the ancilla. The
state of the ancilla in an $n$-fold tensor product of the form $\rho^E=
\rho^e\otimes \cdots \otimes\rho^e$. In this particular case we can use the
quantum coding results~\cite{Sch95,MHor98} to estimate $r$ in the limit of
large $n$; $r=n S(\rho^e)=n S(\rho)$ qubits, where $S(\rho)$ is the von
Neumann entropy of $\rho$; $S(\rho)=-\tr(\rho\log\rho)$.

In the qubit case, the mutual information can be written as
$I(X;Y)=n(1-h(\epsilon))$, where
$h(\epsilon)=-\epsilon\log\epsilon-(1-\epsilon)\log(1-\epsilon)$ is
the binary entropy function and $\epsilon$ is the average bit error
rate. The threshold error rate can be then established from the
condition
\begin{equation}
1-h(\epsilon)\ge S(\rho).
\end{equation}

A key distribution protocol should allow Alice and Bob to estimate the purity
of the pairs of quantum states in terms of the von Neumann entropy $S(\rho)$.
If not they need to maximize $S(\rho)$ over all possible density operators
$\rho$ which are consistent with the estimated bit error rate $\epsilon$.

Moreover the key rate $R$ is
\begin{equation}
  R = H(X) - H(X|Y) - \max_{\rhoh \in \cR} S(\rhoh) \ ,
\end{equation}
(see section \ref{subsubsection:PA})

The argument above relies on the extension of the applicability of
classical privacy amplification to the cases where Eve has partial
quantum rather than partial classical information about the key. This
follows from a more general observation that encoding classical
information into qubits rather than bits, although never worse, does
not offer any significant advantage in some scenarios; ours being one
of them. Amazingly enough potential advantages of quantum encoding
were already pointed out by Wiesner in his seminal paper on conjugate
coding~\cite{Wie70}. Subsequently Ambainis, Nayak, Ta-Shma and
Vazirani~\cite{ANTSU99} considered a scenario where one has to use
partial information about a binary string $X$ to answer a random
binary question about $X$. One might think that storing partial
information about $X$ in a quantum rather than classical memory has a
natural advantage because one can delay a measurement on the quantum
memory until after the question has been asked. This gives an extra
freedom of choosing the most appropriate measurement. However,
Ambainis {\it et al.}~\cite{ANTSU99} and Nayak~\cite{N99} showed that
if information about an $n$--bit string $X$ is stored in $r$ qubits
and one is asked about a particular bit of $X$ then in order to err
with probability less than $\epsilon$ one needs $r>n\;
(1-h(\epsilon))$. Thus, asymptotically, in this particular case,
quantum storage does not offer any advantage. K\"onig, Maurer and
Renner~\cite{KMR03} show that there is no advantage even if one is
asked more general, non-binary, questions about $X$. This made it
possible to make the connection to privacy amplification.

In the following we provide a detailed and reasonably self-contained
description of the new security proof. In section~\ref{sec:preliminaries} we
introduce the relevant concepts and methods of probability theory and quantum
mechanics. The main results are presented in section~\ref{sec:main}. This is
followed by applications of our security criteria to selected quantum key
distribution protocols (section~\ref{sec:examples}).

\section{Preliminaries}
\label{sec:preliminaries}

\subsection{Notation}

Let $\setA$ be a subset of a set $\cI$.  The \emph{characteristic
  function} $\chi_{\setA}$ of $\setA$ on $\setI$ is the function from
$\setI$ to $\{0,1\}$ defined by $\chi(i) = 1$ if and only if $i \in \setA$.

Let $\bz=(z_1, \ldots, z_n)$ be an $n$-tuple and $\setA \subseteq \{1,
\ldots, n\}$ a set of indices. Then $\bz_{\setA}$ denotes the $|\setA|$-tuple
containing all $z_i$ with $i \in \setA$. For two $n$-tuples $\bz=(z_1,
\ldots, z_n)$ and $\bzp=(\zp_1, \ldots, \zp_n)$ of real values, $\bz$ is said
to be \emph{majorized} by $\bzp$, denoted $\bz \prec \bzp$, if for any $k \in
\{1, \ldots, n\}$
\[
\sum_{i \in \setA_k} z_i \leq \sum_{j \in \setB_k} \zp_j
\]
where $\setA_k$ and $\setB_k$ are the sets containing the indices of the $k$
largest elements of $\bz$ and $\bzp$, respectively. A real valued function
$f$ on the set of real $n$-tuples is said to be \emph{Schur-convex} if
\[
  \bz \prec \bzp \implies f(\bz) \leq f(\bzp)
\]
for any $\bz$ and $\bzp$.

For a function $f$ on $\cZ$, we denote by $\fmax{f}$ and $\fmin{f}$ the
functions on the power set of $\cZ$ defined by
\[
  \fmax{f}(\cW) = \max_{z \in \cW} f(z)
\quad \text{and} \quad
  \fmin{f}(\cW) = \min_{z \in \cW} f(z) \ ,
\]
for any $\cW \subseteq \cZ$.

Let $\dist: \cZ \times \cZ \rightarrow \bbR^+$ be a metric on a set $\cZ$.
The \emph{$\eps$-environment of an element} $z \in \cZ$ is defined by
\[
  \cB^{\eps}(z) := \{z' \in \cZ : \dist(z, z') \leq \eps \} \ .
\]
Similarly, the \emph{$\eps$-environment of a subset} $\cW \subseteq
\cZ$ is the union of all $\eps$-environments of elements of $\cW$,
i.e.,
\[
\cB^{\eps}(\cW) := \bigcup_{z \in \cW} \cB^{\eps}(z) \ .
\]

\subsection{Elements of Classical Probability and Information Theory}

The goal of this subsection is to introduce some concepts of
probability and information theory that we will use for the proofs of
our main results. For a more complete overview, we refer to the
standard literature (e.g., \cite{CovTho91}).

 In the following, we use capital letters
($Z$) for random variables, calligraphic letters ($\cZ$) for their
range, and small letters ($z$) for the elements of their range.  The
probability distribution of a random variable $Z$ is denoted by $P_Z$.
The \emph{expectation over $Z$} of a function $f$ of $\cZ$ is given by
$E_Z[f(Z)] := \sum_{z \in \cZ} P_Z(z) f(z)$. A random variable or
probability distribution is called \emph{binary} if it has range $\cZ
= \{0,1\}$.  We write $\Pbin{p}$ for the binary probability
distribution with $\Pbin{p}(1)=p$.

An $n$-tuple $(Z_1, \ldots, Z_n)$ of random variables with the same range
$\cZ$ is called \emph{exchangeable} if, for all permutations $\pi$ on $\{1,
\ldots, n\}$,
\[
  P_{Z_1 \cdots Z_n} = P_{Z_{\pi(1)} \cdots Z_{\pi(n)}}
\]
It is easy to see that, for any $n$-tuple $(Z_1, \ldots, Z_n)$ of random
variables with range $\cZ$, the $n$-tuple
\[
(Z'_1, \ldots, Z'_n) := (Z_{\Pi(1)}, \ldots, Z_{\Pi(n)})
\]
obtained by permuting the indices according to a random permutation $\Pi$ on
$\{1, \ldots, n\}$ is exchangeable.

The \emph{variational distance} between two probability distributions $P$ and
$\Pp$ over the same range $\cZ$ is defined by
\[
  \dist(P,\Pp) := \frac{1}{2} \sum_{z \in \cZ} |P(z) - \Pp(z)| \ .
\]
The variational distance $\dist$ is a metric on the set of probability
distributions with range $\cZ$. In particular, $\dist(P, \Pp) = 0$ if and
only if $P = \Pp$, it is symmetric, and it satisfies the triangle inequality.
For random variables $Z$ and $Z'$, we also write $\dist(Z,Z')$ instead of
$\dist(P_Z, P_{Z'})$. The variational distance between two probability
distributions $P$ and $\Pp$ can be interpreted as the probability that two
random experiments described by $P$ and $\Pp$, respectively, are different.
This is formalized by the following lemma.

\begin{lemma} \label{lem:vardistevent}
  Let $P$ and $\Pp$ be two probability distributions.  Then there
  exists a pair of random variables $Z$ and $\Zp$ with joint
  probability distribution $P_{Z \Zp}$ such that $P_Z = P$, $P_{\Zp} =
  \Pp$, and
  \[
    \Prob[Z \neq \Zp] = \dist(P, \Pp) \ .
  \]
\end{lemma}

It is easy to see that the variational distance between $Z$ and $Z'$ can not
increase when applying the same function $f$ on both $Z$ and $Z'$, i.e.,
\begin{equation} \label{eq:distf}
  \dist(Z,Z') \geq \dist(f(Z),f(Z')) \ .
\end{equation}
Let $[Z,W]$ and $[Z',W']$ be two pairs of random variables, and let
$P_{Z|W}(\cdot, w):=P_{Z|W=w}$ and $P_{Z'|W'}(\cdot, w):=P_{Z'|W'=w}$ be the
probability distribution of $Z$ and $Z'$ conditioned on $W=w$ and $W'=w$,
respectively.  Using the triangle inequality, it can be shown that
\begin{equation} \label{eq:vardistexpgen}
    \bigl| \dist(P_{Z W}, P_{Z',W'})
    - E_W[\dist(P_{Z|W}(\cdot,W), P_{Z'|W'}(\cdot, W))] \bigr|
  \leq
    \dist(P_W,P_{W'}) \ .
\end{equation}
Combining this with~(\ref{eq:distf}) for the function $f: (z,w) \mapsto w$
leads to
\begin{equation} \label{eq:vardistexpbound}
    E_W[\dist(P_{Z|W}(\cdot, W), P_{Z'|W'}(\cdot, W))]
  \leq
    2 \dist(P_{Z W}, P_{Z' W'}) \ ,
\end{equation}
and, similarly, for $f: (z,w) \mapsto z$,
\begin{equation} \label{eq:vardistconv}
    \dist(P_Z,P_{Z'})
  \leq
    E_W[\dist(P_{Z|W}(\cdot,W), P_{Z'|W'}(\cdot, W))] + \dist(P_W,P_{W'}) \ .
\end{equation}

Let $P$ be a probability distribution over $\cZ$. The \emph{non-uniformity}
of $P$,
\[
  d(P) := \dist(P,U) \ ,
\]
is defined as the variational distance of $P$ from the uniform distribution
$U$ over $\cZ$. For a random variable $Z$ with probability distribution
$P_Z$, we also write $d(Z)$ instead of $d(P_Z)$. Similarly, for two random
variables $Z$ and $W$, the \emph{expected non-uniformity} of $Z$ given $W$ is
defined by
\[
  d(Z|W):=E[d(P_{Z|W}(\cdot, W))] \ .
\]

\begin{definition}
  Let $\bz:=(z_1, \ldots, z_n)$ be an $n$-tuple of elements from a set
  $\cZ$. The \emph{frequency distribution} $\freq{\bz}$ of $\bz$ is the
  real valued function on $\cZ$ defined by
  \[
    \freq{\bz}(z) := \frac{|\{i: z_i = z\}|}{n}
  \]
  for $z \in \cZ$.
\end{definition}

\noindent It is easy to see that the frequency $\freq{\bz}$ is a probability
distribution on $\cZ$, i.e., $\freq{\bz}(z) \in [0,1]$ and $\sum_{z \in
  \bZ} \freq{\bz}(z) = 1$.

\begin{definition} \label{def:probrange}
  The \emph{probability range} of an $n$-tuple $\bZ = (Z_1, \ldots,
  Z_n)$ of random variables with range $\cZ$ is the smallest convex
  set $\cP$ of probability distributions on $\cZ$ such that
  \[
    P_{Z_{k}|Z_1=z_1, \ldots, Z_{k-1}=z_{k-1}} \in \cP
  \]
  for all $k \in \{1, \ldots, n\}$ and $z_1, \ldots, z_{n-1} \in \cZ$.
\end{definition}

The following result of~\cite{HolRen03u} states that the frequency
distribution of a sequence of random variables is with high
probability contained in an $\eps$-environment of its probability
range.

\begin{lemma} \label{lem:freq}
  Let $\bZ=(Z_1, \ldots, Z_n)$ be an $n$-tuple of random variables
  with alphabet $\cZ$ of size $|\cZ| = q$ and let $\cP$ be the
  probability range of $\bZ$. Then, for any $\eps>0$,
  \[
    \Prob[\freq{\bZ} \in \cB^{\eps}(\cP)]
  \geq
    1 - 2^q e^{-n \eps^2 / 2} \ .
  \]
\end{lemma}

We will make use of different entropy measures to characterize random
variables or, more precisely, their probability distributions.  Let
$P$ be a probability distribution with range $\cZ$, support
$\cZ^+:=\{z \in \cZ: {P(z)>0}\}$, and maximum probability
$p_{\max}(P):=\max_{z \in \cZ} P(z)$.  Then, the \emph{R\'enyi entropy
  of order $\alpha$}, for $\alpha \in \bbR^+ \cup
\{\infty\}$,\footnote{For $\alpha \in \{0,1,\infty\}$, $H_\alpha(P)$
  is defined by the limit value $\lim_{\beta \to \alpha}
  H_{\beta}(P)$.} is defined by\footnote{All logarithms in this paper
  are binary.}
\[
  H_{\alpha}(P)
:=
  \frac{1}{1-\alpha} \log \bigl(\sum_{z \in \cZ} P(z)^\alpha \bigr) \ .
\]
It turns out that, for $\alpha=1$, $H_1(P)$ corresponds to the \emph{Shannon
entropy} $H(P) = -\sum_{z \in \cZ^+} P(z) \log(P(z))$. Moreover, for
$\alpha=\infty$, we have $H_{\infty}(P) = -\log(p_{\max})$, which is also
called \emph{min-entropy}, and, for $\alpha = 0$, $H_0(P)=\log(|\cZ^+|)$. For
a random variable $Z$ with probability distribution $P_Z$, we also write
$H(Z)$ instead of $H(P_Z)$, and, more generally, for an event $\event$,
$H(Z|\event)$ instead of $H(P_{Z|\event})$.

The R\'enyi entropy of order $\alpha$ of a random variable $Z$ conditioned on
another random variable $W$ is given by
\[
  H_{\alpha}(Z|W)
:=
  \min_{w \in \cW} H_{\alpha}(Z|W=w) \quad \text{(for $\alpha>1$)}
\]
and
\[
  H_{\alpha}(Z|W)
:=
  \max_{w \in \cW} H_{\alpha}(Z|W=w) \quad \text{(for $\alpha<1$)} \ .
\]

We will often be interested in the entropy of a probability
distribution which is close to a given distribution $P$. This is
formalized by the notion of \emph{smooth R\'enyi entropy} introduced
in~\cite{RenWol03u}.

\begin{definition} \label{def:smoothR}
  Let $\eps \geq 0$ and $\alpha \in \cR^+ \cup \{\infty\}$. The
  \emph{$\eps$-smooth R\'enyi entropy or order $\alpha$} of a
  probability distribution $P$ is defined by
  \[  H^{\eps}_{\alpha}(P)  := \fmax{H_{\alpha}}(\cB^{\eps}(P)) \quad \text{(for
  $\alpha>1$)}\]
    and
  \[  H^{\eps}_{\alpha}(P) := \fmin{H_{\alpha}}(\cB^{\eps}(P)) \quad \text{(for $\alpha<1$)} \ .\]
\end{definition}

Similarly, the notion of conditional R\'enyi entropy can be generalized to
smooth R\'enyi entropy. In particular, for $\alpha=\infty$, we have
\[
  H_{\infty}^{\eps}(Z|W)
:=
  \max_{P_{Z' W'}: \dist(P_{Z' W'}, P_{Z W})\leq \eps} H_{\infty}(Z'|W') \ .
\]


\noindent The following lemma is an immediate consequence of the above definition
for $\alpha = 0$.

\begin{lemma} \label{lem:H0epscomp}
  Let $Z$ be a random variable with range $\cZ$ and let $\cW$ be a
  subset of $\cZ$. Then, for any $\eps \geq 0$,
  \[
    \Prob[Z \in \cW] \geq 1-\eps
  \quad \implies \quad
    H_0^{\eps}(P_Z) \leq \log |\cW| \ .
  \]
\end{lemma}

For $\alpha=0$, the (smooth) R\'enyi entropy is sub-additive.

\begin{lemma} \label{lem:entradd}
  Let $Z$ and $W$ be two random variables. Then, for any $\eps,
  \epsp>0$,
  \[
    H_0^{\eps+\epsp}(Z W) \leq H_0^{\eps}(Z) + H_0^{\epsp}(W) \ .
  \]
\end{lemma}

The min-entropy of a random variable $Z$ when conditioning on another random
variable $W$ cannot decrease more than the R\'enyi entropy of order zero of
$W$.

\begin{lemma} \label{lem:Hinfsub}
  Let $Z$ and $W$ be random variables. Then, for any $\eps, \epsp,
  \epspp \in \bbR^+$,
  \[
    H_{\infty}^{\eps + \epsp + \epspp}(Z|W)
  \geq
    H_{\infty}^{\eps}(Z W) - H_0^{\epsp}(W) - \log(\frac{1}{\epspp}) \ .
  \]
\end{lemma}

\begin{lemma} \label{lem:Hinfex}
  Let $\bZ$ be an exchangeable $n$-tuple of random variables with
  range $\cZ$. Then
  \[
    H_{\infty}(\bZ|Q_{\bZ}=Q)
  \geq
    n H(Q) - |\cZ| (\log(n) + 1) \ .
  \]
\end{lemma}

\begin{proof}
  By the definition of exchangeability, $P_{\bZ|Q_{\bZ}=Q}$ is the
  uniform distribution over the set of all $n$-tuples $\bz$ with
  $Q_{\bz} = Q$. It is easy to see that there are
  \[
    N_Q := \frac{n!}{\prod_{z \in \cZ} (n Q(z))!}
  \]
  such tuples, i.e., we have $H_{\infty}(\bZ|Q_{\bZ}=Q) = - \log
  p_{\max} (P_{\bZ|Q_{\bZ}=Q}) = \log(N_Q)$. The assertion then follows
  from a straightforward calculation using Stirling's approximation
  \[
     \sqrt{2 \pi} m^{m+\frac{1}{2}} e^{-m}
  \leq
    m!
  \leq
    m^{m+\frac{1}{2}} e^{-m+1} \ ,
  \]
  for any $m \in \bbN$.
\end{proof}

The notion of typical sets is widely used in information theory.  Note
that the following definition slightly differs from the one given
in~\cite{CovTho91}.

\begin{definition} \label{def:typset}
  Let $\cZ$ be a set, $n \in \bbN$, and $r \geq 0$.  The
  \emph{$r$-typical set over $\cZ^n$} is defined as
  \[
  \cT^n_{\cZ}(r) := \{\bz \in \cZ^n : H(\freq{\bz}) \leq r \} \ .
  \]
\end{definition}

\begin{lemma} \label{lem:typsetsize}
  For any set $\cZ$ of size $|\cZ|=q$, $n, \in \bbN$, and $r \geq 0$,
  \[
    |\cT^n_{\cZ}(r)| \leq 2^{n r} n^{q-1} \ .
  \]
\end{lemma}

\begin{proof}
  Let $\cQ:=\{\freq{\bz}: \bz \in \cZ^n\}$ be the set of frequency
  distributions of $n$-tuples over $\cZ$ and, for any $\Qh \in \cQ$,
  let $\cS(\Qh):= \{\bz \in \cZ^n : \freq{\bz} = \Qh \}$ be the set of
  $n$-tuples $\bz=(z_1, \ldots, z_n)$ with frequency distribution
  $\Qh$.  We first show that, for any $\Qh \in \cQ$,
  \begin{equation} \label{eq:SQbound}
    |\cS(\Qh)| \leq 2^{n H(\Qh)} \ .
  \end{equation}
  Let $\bZ = (Z_1, \ldots, Z_n)$ be an $n$-tuple of independent random
  variables $Z_i$ distributed according to $\Qh$. Since, for any
  $n$-tuple $\bz$ in $\cS(\Qh)$, each symbol $z$ occurs $n \Qh(z)$
  times in $\bz$, we find
  \[
    1
  \geq
    \Prob[\bZ \in \cS(\Qh)]
  =
    |\cS(\Qh)| \prod_{z \in \cZ} \Qh(z)^{n\Qh(z)}
  =
    |\cS(\Qh)| 2^{\sum_{z \in \cZ} n \Qh(z) \log(\Qh(z))}
  \]
  which implies~(\ref{eq:SQbound}).  The assertion of the lemma then
  follows from
  \[
    |\cT^n_{\cZ}(r)|
  =
    \sum_{\Qh \in \cQ: \, H(\Qh) \leq r} |\cS(\Qh)|
  \leq
    \sum_{\Qh \in \cQ: \, H(\Qh) \leq r} 2^{n H(\Qh)}
  \leq
    |\cQ| \, 2^{n r}
  \]
  and the observation that $|\cQ| \leq n^{q-1}$.
\end{proof}

\begin{definition}
  Let $p \in [0,1]$ and let $\setI$ be a set. A \emph{$p$-random
    selection} $A$ on $\cI$ is a random variable describing the subset
  obtained by independently picking each element of $\setI$ with
  probability $p$, i.e., for any $\setA \subseteq \setI$,
  \[
    P_A(a) = \prod_{i \in \setI} \Pbin{p}(\chi_{\setA}(i))
  \]
  where $\chi_{\setA}$ be the characteristic function of $\setA$ on
  $\setI$.
\end{definition}

A random function $G$ from $\cX$ to $\cY$ is called \emph{\tu{}} if $
\Prob[G(x) = G(x')] \leq \frac{1}{|\cY|} $ holds for any distinct $x, x' \in
\cX$. In particular, $G$ is \tu{} if, for any distinct $x, x' \in \cX$, the
random variables $G(x)$ and $G(x')$ are independent and uniformly
distributed. For instance, the uniform random function from a set $\cX$ to a
set $\cY$ is \tu.\footnote{In the literature, \tuty{}
  is usually defined for families $\cG$ of functions: A family $\cG$
  is called \tu{} if the random function $G$ with uniform distribution
  over $\cG$ is \tu. Non-trivial examples of \tu{} families $\cG$ of
  functions can, e.g., be found in~\cite{CarWeg79} and~\cite{WC81}.}

\subsection{Elements of Quantum Theory}

In this section, we introduce some basic concepts of quantum theory
which we will use.  For a more complete overview we refer to the
standard literature (e.g., \cite{NieChu00}).

Let $\cH$ be a Hilbert space of dimension $d$. We denote by $\cS(\cH)$
the set of \emph{density operators} on $\cH$, i.e., $\cS(\cH)$ is the
set of positive operators $\rho$ on $\cH$ with $\tr(\rho)=1$. For any
$\rho \in \cS(\cH)$, let $\ev(\rho)$ be the $d$-tuple of eigenvalues
of $\rho$ (e.g., in decreasing order). The \emph{trace distance}
between two density operators $\rho$ and $\sigma$ on the same Hilbert
space $\cH$ is defined by
\[
  \dist(\rho, \sigma) := \frac{1}{2} \tr(|\rho-\sigma|) \ .
\]
We will use several well-known properties of the trace distance (for
proofs, see e.g. \cite{NieChu00}).

The trace distance can be seen as a generalization of the variational
distance to density operators. Many of the properties of the variational
distance thus also hold for the trace distance. In particular, the trace
distance is a metric on $\cS(\cH)$.

Moreover, for two probability distributions $P$ and $\Pp$ over $\cW$ and two
families of density operators $\{\rho_w\}_{w \in \cW}$ and $\{\sigma_w\}_{w
\in \cW}$,
\begin{equation} \label{eq:trdistconv}
  \dist(\sum_{w \in \cW} P(w) \rho_w, \sum_{w \in \cW} \Pp(w) \sigma_w)
\leq
  \sum_{w \in \cW} P(w) \, \dist(\rho_w, \sigma_w) + \dist(P, \Pp) \ .
\end{equation}
This inequality can be seen as the quantum analogue
of~(\ref{eq:vardistconv}).  The trace distance between two \emph{pure
  states} $\rho = \ket{\phi} \bra{\phi}$ and $\sigma = \ket{\psi}
\bra{\psi}$ can easily be computed explicitly,
\begin{equation} \label{eq:trdistpure}
    \dist(\rho, \sigma)
  =
    \sqrt{1-|\spr{\phi}{\psi}|^2}
  =
    \sqrt{1-\tr(\rho \sigma)} \ .
\end{equation}

Let $\cF$ be a \emph{positive operator valued measure (POVM)} on a Hilbert
space $\cH$, i.e., $\cF= \{F_z\}_{z \in \cZ}$ is a family of positive
operators on $\cH$ such that $\sum_{z \in \cZ} F_z = \id$. We say that $\cF$
is \emph{orthogonal} if there exists an orthonormal basis $\{\ket{z}\}_{z \in
\cZ}$ of $\cH$ such that $F_z = \ket{z}\bra{z}$, for any $z \in \cZ$.

\begin{definition}
  Let $\cF = \{F_z\}_{z \in \cZ}$ be a POVM on a Hilbert space $\cH$.
  The \emph{measurement mapping} $\meas_{\cF}$ of $\cF$ is the
  function mapping each density operator $\rho \in \cS(\cH)$ to the
  probability distribution $P = \meas_{\cF}(\rho)$ on $\cZ$ defined by
  $P(z) := \tr(F_z \rho)$.  The \emph{probability range} $\cP_{\cF}$
  of $\cF$ is the range of $\meas_{\cF}$, i.e., $\cP_{\cF}:=
  \meas_{\cF}(\cS(\cH))$.
\end{definition}

For a POVM $\cF = \{F_z\}_{z \in \cZ}$ on a Hilbert space $\cH$ and a
probability distribution $P$ on $\cZ$, we write $\meas_{\cF}^{-1}(P)$ to
denote the set of density operators $\rho$ on $\cH$ such that
$\meas_{\cF}(\rho) = P$. More generally, for a set $\cP$ of probability
distributions, $\meas_{\cF}^{-1}(\cP):=\bigcup_{P \in \cP}
\meas_{\cF}^{-1}(P)$ is the set of density operators $\rho$ with
$\meas_{\cF}(\rho) \in \cP$

The trace distance between two density operators $\rho$ and $\sigma$ turns
out to be an upper bound for the variational distance between the probability
distributions of the outcomes of the same measurement $\cF$ applied to $\rho$
and $\sigma$.

\begin{lemma} \label{lem:measdist}
  Let $\cF$ be a POVM on a Hilbert space $\cH$ and let $\rho, \sigma
  \in \cS(\cH)$. Then
  \[
    \dist(\meas_{\cF}(\rho), \meas_{\cF}(\sigma)) \leq \dist(\rho, \sigma) \ .
  \]
\end{lemma}

The probability distribution resulting from an orthogonal measurement of a
quantum state $\rho$ is in a certain sense less ordered than the eigenvalues
of $\rho$. This is formalized by the following lemma. A proof can, for
instance, be found in~\cite{HorJoh85} (see also~\cite{Bhatia96}).

\begin{lemma}[Schur's majorization theorem] \label{lem:Schur}
  Let $\cF=\{F_1, \ldots, F_d\}$ be an orthogonal measurement on a
  $d$-dimensional Hilbert space. Then, for any density operator $\rho
  \in \cS(\cH)$,
  \[
    \bp \prec \ev(\rho)
  \]
  where $\bp=(\meas_{\cF}(\rho)(1), \ldots, \meas_{\cF}(\rho)(d))$ are
  the probabilities of the outcomes when measuring $\rho$ with respect
  to $\cF$.
\end{lemma}

Let $\cH \otimes \cH'$ be a bipartite Hilbert space, let $\rho \in
\cS(\cH \otimes \cH')$, and let $Z$ be the outcome of a measurement of
$\rho$ with respect to a POVM $\cF = \{F_z\}_{z \in \cZ}$ on (a
subspace of) $\cH'$. The density operator on $\cH$ resulting from
conditioning $\rho$ on the measurement outcome $Z=z$, denoted
$\rho^{\cH}_{z \leftarrow \cF}$, is given by
\[
  \rho^{\cH}_{z \leftarrow \cF}
:=
  \frac{1}{c} \tr_{\cH'}(\id_{\cH} \otimes F_z \, \rho)
\]
where $c := \tr(\id_{\cH} \otimes F_z \, \rho)$ is a normalization
constant and where $\tr_{\cH'}$ denotes the partial trace over the
subspace $\cH'$.

Let $\cH^{\otimes n}:=\cH_1 \otimes \cdots \otimes \cH_n$ be the product of
$n$ identical factor spaces $\cH_i = \cH$.  The following definition can be
seen as a quantum version of Definition~\ref{def:probrange}.

\begin{definition}
  The \emph{density range} of a density operator $\rho \in
  \cH^{\otimes n}$ is the smallest convex subset $\cP$ of $\cS(\cH)$
  such that for any $k \in \{1, \ldots, n\}$, for any POVM $\cF^{k-1}
  = \{F_z\}_{z \in \cZ}$ on $\cH^{k-1}:=\bigotimes_{i=1}^{k-1} \cH_i$,
  and for any $z \in \cZ$, the density operator $\rho^{\cH_k}_{z
    \leftarrow \cF^{k-1}}$ is contained in $\cP$.
\end{definition}

Let $\cE$ be a \emph{quantum operation} $\cE$ on a Hilbert space $\cH$, i.e.,
$\cE = \{E_z\}_{z \in \cZ}$ is a family of linear operators $E_z$ on $\cH$
such that $\sum_{z \in \cZ} E_z^{\dagger} E_z = \id$. Then, the density
operator $\sigma = \cE(\rho)$ \emph{resulting from applying} $\cE$ to a
density operator $\rho$ is given by
\[
  \sigma := \sum_{z \in \cZ} E_z^{\vphantom{\dagger}} \rho E_z^{\dagger} \ .
\]

\begin{lemma} \label{lem:proj}
  Let $\rho$ be a density operator on $\cH$ and let $\sigma :=
  \cE(\rho)$ be the density operator resulting from applying a quantum
  operation $\cE = \{E_z\}_{z \in \cZ}$ to $\rho$. Then
  \[
    \dist(\rho, \sigma) \leq \sqrt{1 - \sum_{z \in \cZ} |\tr(E_z \rho)|^2} \ .
  \]
\end{lemma}

\begin{proof}
  We first show that the assertion of the lemma holds if $\rho =
  \ket{\phi}\bra{\phi}$ is a pure state. For $z \in \cZ$, let $p_z :=
  \tr(E_z \rho E^{\dagger}_z)$, $\ket{\psi_z} := \frac{1}{\sqrt{p_z}}
  E_z \ket{\phi}$, and $\sigma_z:=\ket{\psi_z}\bra{\psi_z}$.  Note
  that $p_z \in [0,1]$, $\sum_z p_z = 1$, and
  \[
    \sigma = \sum_{z \in \cZ} p_z \sigma_z \ .
  \]
  We can thus apply~(\ref{eq:trdistconv}) yielding
  \begin{equation} \label{eq:distsum}
    \dist(\rho, \sigma)
  \leq
    \sum_{z \in \cZ} p_z \dist(\rho, \sigma_z) \ .
  \end{equation}
  Since, $\rho = \ket{\phi}\bra{\phi}$ and
  $\sigma_z=\ket{\psi_z}\bra{\psi_z}$ are pure states, it follows
  from~(\ref{eq:trdistpure}) that
  \[
    \dist(\rho, \sigma_z)
  =
    \sqrt{1-|\spr{\phi}{\psi_z}|^2}
  =
    \sqrt{1-\frac{1}{p_z} |\tr(E_z \rho)|^2} \ .
  \]
  Combining this with~(\ref{eq:distsum}), we find
  \[
    \dist(\rho, \sigma)
  \leq
    \sum_{z \in \cZ} p_z \sqrt{1-\frac{1}{p_z} |\tr(E_z \rho)|^2}
  \leq
    \sqrt{\sum_{z \in \cZ} p_z \bigl(1-\frac{1}{p_z} |\tr(E_z \rho)|^2\bigr)}
  \]
  where the second inequality follows from the concavity of the square
  root and Jensen's inequality. This concludes the proof of the lemma
  for pure states $\rho$.

  To verify that the assertion of the lemma also holds for mixed
  states $\rho$, write $\rho$ as a convex combination of pure states
  $\rho_w$, i.e., $\rho = \sum_{w \in \cW} q_w \rho_w$ for appropriate
  $q_w \in [0,1]$ with $\sum_{w \in \cW} q_w = 1$, and let
  $\cE(\rho_w)$ be the state resulting from applying the quantum
  operation $\cE$ on $\rho_w$. Then, since $\sigma = \cE(\rho) =
  \sum_{w \in \cW} q_w \cE(\rho_w)$, inequality~(\ref{eq:trdistconv})
  yields
  \[
    \dist(\rho, \sigma)
  \leq
    \sum_{w \in \cW} q_w \dist(\rho_w, \cE(\rho_w))
  \leq
    \sum_{w \in \cW} q_w \sqrt{1 - \sum_{z \in \cZ} |\tr(E_z \rho_w)|^2}
  \]
  where the last inequality follows from the statement of the lemma
  applied to the pure states $\rho_w$. Using again Jensen's
  inequality, we obtain
  \[
    \dist(\rho, \sigma)
  \leq
    \sqrt{1 - \sum_{z \in \cZ} \sum_{w \in \cW} q_w |\tr(E_z \rho_w)|^2}
  \leq
    \sqrt{1 - \sum_{z \in \cZ} |\tr(E_z \sum_{w \in \cW} q_w \rho_w)|^2}
  \]
  which concludes the proof.
\end{proof}

We will now use Lemma~\ref{lem:proj} to derive a lower bound for the
variational distance between two probability distributions in terms of the
trace distance between two corresponding density operators. This is in a
certain sense the converse of Lemma~\ref{lem:measdist}.

\begin{lemma} \label{lem:existdist}
  Let $\cF = \{F_z\}_{z \in \cZ}$ be an orthogonal POVM on a Hilbert
  space $\cH$, let $\rho \in \cS(\cH)$, let $P:=\meas_{\cF}(\rho)$,
  and let $\Pp$ be a probability distribution on $\cZ$. Then there
  exists $\sigma \in \cS(\cH)$ such that
  \begin{equation} \label{eq:condmeas}
    \Pp = \meas_{\cF}(\sigma)
  \end{equation}
  and
  \begin{equation} \label{eq:conddist}
    \dist(\rho, \sigma)
  \leq
    \sqrt{2 \dist(P, \Pp) }  \ .
  \end{equation}
  In particular,
  \begin{equation}
    \cB^{\eps}(\meas_{\cF}(\rho))
  \subseteq
    \meas_{\cF}(\cB^{\sqrt{2 \eps}}(\rho)) \ .
  \end{equation}
\end{lemma}

\begin{proof}
  From Lemma~\ref{lem:vardistevent}, there exist random variables $Z$
  and $\Zp$ distributed according to $P$ and $\Pp$, respectively, such
  that
  \[
    \Prob[Z \neq \Zp] = \delta := \dist(P, \Pp) \ .
  \]
  Let $\cW:=\{(z,\zp) \in \cZ \times \cZ: z \neq \zp\}$, let
  $p_{z,\zp}:=P_{\Zp|Z}(\zp,z)$, and let $\{\ket{z}\}_{z \in \cZ}$ be
  an orthonormal basis of $\cH$ such that $F_z = \ket{z} \bra{z}$. Let
  \[
    E_0^{\vphantom{\dagger}} := \sum_{z \in \cZ} \sqrt{p_{z,z}} \ket{z}\bra{z} \ ,
  \]
  and, for $(z,\zp) \in \cW$,
  \[
  E_{z,\zp}^{\vphantom{\dagger}} := \sqrt{p_{z,\zp}} \ket{\zp}\bra{z} \ ,
  \]
  be linear operators on $\cH$. It is easy to verify that the family
  $\cE = \{E_0\} \cup \{E_{z,\zp}\}_{(z,\zp) \in \cW}$ is a quantum
  operation, i.e.,
  \[
    E_0^{\dagger} E_0^{\vphantom{\dagger}}
    \; + \! \sum_{(z,\zp) \in \cW}
              E_{z,\zp}^{\dagger} E_{z,\zp}^{\vphantom{\dagger}}
  =
    \id \ .
  \]
  Let
  \[
    \sigma := \cE(\rho)
  =
    E_0^{\vphantom{\dagger}} \rho E_0^{\dagger}
    \; + \! \sum_{(z,\zp) \in \cW}
              E_{z,\zp}^{\vphantom{\dagger}} \rho E_{z,\zp}^{\dagger}
  \]
  be the quantum state resulting from applying $\cE$ to $\rho$. It
  then follows from a straightforward calculation that
  \[
    \meas_{\cF}(\sigma) = P_{\Zp}
  \]
  which implies~(\ref{eq:condmeas}) since $P_{\Zp} = \Pp$. To show
  that also~(\ref{eq:conddist}) holds, we use Lemma~\ref{lem:proj}
  yielding
  \[
    \dist(\rho, \sigma)
  \leq
    \sqrt{1 - |\tr(E_0 \rho)|^2
    - \sum_{(z,\zp) \in \cW} |\tr(E_{z,\zp} \rho)| ^2}
  \leq
    \sqrt{1 - |\tr(E_0 \rho)|^2} \ .
  \]
  Since $P_Z = \meas_{\cF}(\rho)$, we have
  \[
    \tr(E_0 \rho)
  =
    \sum_{z \in \cZ} \sqrt{p_{z,z}} \meas_{\cF}(\rho)(z)
  \geq
    \sum_{z \in \cZ} p_{z,z} P_Z(z)
  =
    \Prob[Z = \Zp]
  =
    1- \delta \ ,
  \]
  and thus
  \[
    \dist(\rho, \sigma)
  \leq
    \sqrt{1 - (1-\delta)^2}
  =
    \sqrt{2 \delta - \delta^2}
  \leq
    \sqrt{2 \delta} \ .
  \]
\end{proof}

The entropy of a quantum state can be defined in terms of the entropy
of a classical probability distribution. Let $\rho$ be a density
operator on a $d$-dimensional Hilbert space $\cH$ and let $(\lambda_1,
\ldots, \lambda_d):=\ev(\rho)$ be the $d$ eigenvalues of $\rho$.  Note
that there exists an orthonormal basis $\{\ket{1}, \ldots \ket{d}\}$
of $\rho$ (namely the eigenbasis) such that $\lambda_i = P(i)$ where
$P := \meas_{\cF}(\rho)$ is the probability distribution of a
measurement of $\rho$ with respect to the POVM $\cF =
\{\ket{1}\bra{1}, \ldots, \ket{d}\bra{d}\}$.  In particular
$\ev(\rho)$ can be interpreted as a probability distribution on $\{1,
\ldots, d\}$.

The \emph{R\'enyi entropy} (of order $\alpha$) of a density operator $\rho$
is defined by the R\'enyi entropy of $\ev(\rho)$, i.e., $\qH_{\alpha}(\rho)
:= H_{\alpha}(\ev(\rho))$, for $\alpha \in \bbR^+ \cup \{\infty\}$. In
particular, for $\alpha=1$, $\qH(\rho) := \qH_1(\rho)$ is the \emph{von
Neumann entropy} of $\rho$. Note that, for $\alpha = 0$,
\[
  \qH_0(\rho) = \log(\rank(\rho)) \ .
\]
The smooth R\'enyi entropy for density operators can be defined
by generalizing the classical Definition~\ref{def:smoothR}.

\begin{definition}
  Let $\eps \geq 0$ and $\alpha \in \cR^+ \cup \{\infty\}$. The
  \emph{$\eps$-smooth R\'enyi entropy of order $\alpha$} of a
  density operator $\rho$ is defined by
  \[
    \qH_{\alpha}(\rho)
  :=
    \fmax{\qH_{\alpha}}(\cB^{\eps}(\rho)) \quad \text{(for $\alpha>1$)}
  \quad \text{and} \quad
    \qH_{\alpha}(\rho)
  :=
    \fmin{\qH_{\alpha}}(\cB^{\eps}(\rho)) \quad \text{(for $\alpha<1$)} \ .
  \]
\end{definition}

The following lemma is a direct consequence of Lemma~\ref{lem:Schur} and the
fact that the entropy functions $-H_\alpha$ are Schur-convex.

\begin{lemma} \label{lem:Halphadec}
  Let $\cF$ be an orthogonal POVM on a $d$-dimensional Hilbert space.
  Then, for any density operator $\rho \in \cS(\cH)$ and any $\alpha
  \in \bbR^+ \cup \{\infty\}$,
  \[
    \qH_\alpha(\rho) \leq H_\alpha(\meas_{\cF}(\rho)) \ .
  \]
\end{lemma}

We often will use this result for the case $\alpha=1$. To simplify the
notation, let
\begin{equation} \label{eq:POVMentropy}
  \qH_{\cF}(\rho) := H(\meas_{\cF}(\rho)) .
\end{equation}
be the Shannon entropy of the outcomes when measuring a density operator
$\rho \in \cS(\cH)$ with respect to a POVM $\cF$. If $\cF$ corresponds to a
measurement in an eigenbasis of $\rho$, we obviously have $\qH(\rho) =
\qH_{\cF}(\rho)$, and thus, from Lemma~\ref{lem:Halphadec},
\begin{equation} \label{eq:entrbasis}
  \qH(\rho) = \min_{\cF} \qH_{\cF}(\rho)
\end{equation}
where the minimum is taken over all orthogonal POVMs $\cF$ in $\cH$.

The following lemma is an extension of Lemma~\ref{lem:Halphadec} to smooth
R\'enyi entropy.

\begin{lemma} \label{lem:H0epsdec}
  Let $\cF$ be an orthogonal POVM on a Hilbert space $\cH$.  Then, for
  any density operator $\rho \in \cS(\cH)$. Then, for any density
  operator $\rho$, $\alpha<1$, and $\eps \geq 0$,
  \[
    \qH_\alpha^{\sqrt{2 \eps}}(\rho)
  \leq
    H_\alpha^{\eps}(\meas_{\cF}(\rho)) \ .
  \]
\end{lemma}

\begin{proof}
  From Lemma~\ref{lem:Halphadec}, we have
  \[
    \qH_\alpha^{\sqrt{2 \eps}}(\rho)
  =
    \inf_{\sigma \in \cB^{\sqrt{2 \eps}}(\rho)} \qH_\alpha(\sigma)
  \leq
    \inf_{\sigma \in \cB^{\sqrt{2 \eps}}(\rho)}
      H_\alpha(\meas_{\cF}(\sigma)) \ .
  \]
  The assertion then follows from Lemma~\ref{lem:existdist},
  \[
    \inf_{\sigma \in \cB^{\sqrt{2 \eps}}(\rho)}
      H_\alpha(\meas_{\cF}(\sigma))
  \leq
    \inf_{\Pp \in \cB^{\eps}(\meas_{\cF}(\rho))} H_\alpha(\Pp)
  =
    H_\alpha^{\eps}(\meas_{\cF}(\rho)) \ .
  \]
\end{proof}

\section{Main Result} \label{sec:main}

This section contains the main result of the paper, namely, an explicit
expression for the rate of secure quantum key distribution (cf.
equation~(\ref{eq:rate})). In the first part, we derive
Lemma~\ref{lem:quanttom} which says that the frequency distribution obtained
when measuring the subsystems of an $n$-partite quantum state with respect to
a certain POVM $\cFb$ can be estimated from the results obtained by applying
another POVM $\cF$ on a few randomly chosen subsystems. This is then used to
show Lemma~\ref{lem:quantentrbasic} which gives an upper bound for the
R\'enyi entropy of order $0$ of the outcomes when applying the POVM $\cF$
given only the outcomes of the measurements with respect to $\cFb$ on a few
(randomly chosen) subsystems. The result is then applied to bound the size
(rank) of the $n$-partite quantum system given the outcomes of a measurement
on a few subsystems (Corollary~\ref{cor:quantentr}).

In Sections~\ref{sec:IR} and~\ref{sec:PA} we review information
reconciliation and the security of privacy amplification in the presence of a
quantum adversary, respectively. These are main ingredients of the
post--processing stage.

In Section~\ref{sec:protocol}, we introduce the generic quantum key
distribution protocol and prove its security by combining the above mentioned
results with the information reconciliation and privacy amplification to
obtain our main result, i.e., the secret key rate~(\ref{eq:rate}).

\subsection{Parameter Estimation}

Let $\cH$ be a Hilbert space, let $\rho \in \cS(\cH^{\otimes n})$, let $a
\subseteq \{1, \ldots, n\}$, and let $\cF = \{F_z\}_{z \in \cZ}$ be a POVM on
$\cH$. Then $\ZVmeas^a_{\cF}(\rho)$ denotes the $|a|$-tuple $\bZh$ of
outcomes resulting from applying $\cF$ to $\rho$ on $\cH_a$, where $\cH_a$ is
the tensor product of the factor spaces $\cH_i$, for $i \in a$.

\label{subsection:tomography}
\begin{lemma} \label{lem:quanttom}
  Let $\rho \in \cS(\cH^{\otimes n})$ be an $n$-partite state with
  density range $\cR \subseteq \cS(\cH)$, let $\cF = \{F_z\}_{z \in
    \cZ}$ and $\cFb = \{\Fb_z\}_{z \in \cZb}$ be two POVMs on $\cH$,
  and let $A$ be a $p$-random selection on $\{1, \ldots, n\}$. Let
  $\bZ_A:=\ZVmeas_{\cF}^A(\rho)$ and
  $\bZ_{\Ab}:=\ZVmeas_{\cFb}^{\Ab}(\rho)$ be the outcomes when
  measuring $\rho$ in $\cH_A$ and $\cH_{\Ab}$ with respect to $\cF$
  and $\cFb$, respectively. Then, for any $\eps>0$,
  \[
    \Prob\bigl[\exists \rhoh \in \cR : \,
      p \, \dist(\freq{\bZ_A}, \meas_{\cF}(\rhoh))
      + (1-p) \, \dist(\freq{\bZ_{\Ab}}, \meas_{\cFb}(\rhoh))
    \leq  \eps \bigr] \geq 1-\mu
  \]
  where $\mu:= 2^{|\cZ|+|\cZb|} e^{-\frac{n \eps^2}{8}}$.
\end{lemma}

\begin{proof}
  Let $\cG$ be the POVM on $\cH$ obtained by combining $\cF$ and
  $\cFb$ with probability $p$ and $1-p$, respectively, i.e., $ \cG :=
  \{G_{(z,r)}\}_{(z,r) \in (\cZ \cup \cZb) \times \{0,1\}} $ with
  $G_{(z,1)} := p F_z$ and $G_{(z,0)} := (1-p) \Fb_z$. Let
  $\bW:=\ZVmeas_{\cG}(\rho)$ be the $n$-tuple of outcomes $(Z_i, R_i)$
  when measuring $\rho$ with respect to $\cG$. The random variables
  occurring in the lemma can then equivalently be defined by $A:=\{i :
  R_i=1\}$, $\bZ_{A}:=(Z_1, \ldots, Z_n)_A$, and $\bZ_{\Ab}:=(Z_1,
  \ldots, Z_n)_{\Ab}$.

  The probability range of the $n$-tuple $\bW$ is contained in
  $\cP:=\meas_{\cG}(\cR)$.  We can thus apply Lemma~\ref{lem:freq} for
  $\epsb := \eps/2$ leading to
  \[
    \Prob[\freq{\bW} \in \cB^{\epsb}(\cP)]
  \geq
    1- \mu \ .
  \]
  Let $\bz \in (\cZ \cup \cZb)^n$ and $\setA \subseteq \{1, \ldots,
  n\}$ such that the $n$-tuple $\bw$ of values $w_i = (z_i,
  \chi_{\setA}(i))$ satisfies $\freq{\bw} \in \cB^{\epsb}(\cP)$, i.e.,
  \[
    \dist(\freq{\bw}, \meas_{\cG}(\rhoh)) \leq \eps/2
  \]
  for some $\rhoh \in \cR$.  It remains to be shown that this implies
  \begin{equation} \label{eq:qQdist}
    p \, \dist(\freq{\bz_a}, \meas_{\cF}(\rhoh))
      + (1-p) \, \dist(\freq{\bz_{\ab}}, \meas_{\cFb}(\rhoh))
  \leq
    \eps \ .
  \end{equation}
  Let $(Z,R)$ and $(\Zp,\Rp)$ be two pairs of random variables
  distributed according to $\meas_{\cG}(\rhoh)$ and $\freq{\bw}$,
  respectively.  It follows from the construction of the POVM $\cG$
  that $P_R = \Pbin{p}$, $P_{Z|R=1} = \meas_{\cF}(\rhoh)$, and
  $P_{Z|R=0} = \meas_{\cFb}(\rhoh)$.  Moreover, by the definition of
  the frequency distribution, $P_{\Zp|\Rp=1} = \freq{\bz_{\setA}}$ and
  $P_{\Zp|\Rp=0} = \freq{\bz_{\setAb}}$.  Hence,
  using~(\ref{eq:vardistexpbound}),
  \[
  \begin{split}
    p \, \dist(\meas_{\cF}(\rhoh), \freq{\bz_{\setA}})
    + (1-p) \, \dist(\meas_{\cFb}(\rhoh), \freq{\bz_{\setAb}})
  & =
    E_R[P_{Z|R}(\cdot, R), P_{\Zp|\Rp}(\cdot, R)] \\
  & \leq
    2 \dist(P_{Z R}, P_{\Zp \Rp})
  =
    2 \dist(\meas_{\cG}(\rhoh), \freq{\bw})
  \leq
    \eps \ .
  \end{split}
  \]
  which implies~(\ref{eq:qQdist}) and thus concludes the proof.
\end{proof}


\begin{lemma} \label{lem:quantentrbasic}
  Let $\rho \in \cS(\cH^{\otimes n})$ be an $n$-partite state with
  density range $\cR \subseteq \cS(\cH)$, let $\cF = \{F_z\}_{z \in
    \cZ}$ and $\cFb = \{\Fb_z\}_{z \in \cZb}$ be two POVMs on $\cH$,
  let $A$ be a $p$-random selection on $\{1, \ldots, n\}$, and let
  $\bZ_A:=\ZVmeas_{\cF}^A(\rho)$,
  $\bZ_{\Ab}:=\ZVmeas_{\cFb}^{\Ab}(\rho)$. Then, for any $\eps>0$,
  there exists a real valued function $\mu$ with
  \begin{equation*} 
    E[\mu(\freq{\bZ_A},A)]
  \leq
    2^{|\cZ|+|\cZ'|} e^{-\frac{n \eps^2}{2}}
  \end{equation*}
  such that, for any probability distribution $\Qh$ on $\cZ$ and any
  $a \subseteq \{1, \ldots, n\}$,
  \[
    H_0^{\mu(\Qh, a)}(\bZ_{\Ab}|\freq{\bZ_A}=\Qh, A=a)
  \leq
    |\ab| \, \fmax{H}(\cB(\Qh)) + \log(|\ab|) (|\cZb|-1) \ ,
  \]
  where $\cB(\Qh):= \cB_{\eps/(1-p)} (\meas_{\cFb}(\cR \cap
  \meas_{\cF}^{-1} (\cB_{\eps / p}(\Qh))))$.
\end{lemma}

\begin{proof}
  Let $\cW$ be the set of pairs $(\bz, a)$ consisting of an $n$-tuple
  $\bz$ of elements from $\cZ \cup \cZb$ and a subset $a \subseteq
  \{1, \ldots, n\}$ such that there exists a density operator $\rhoh
  \in \cR$ satisfying
  \[
    p \, \dist(\freq{\bz_a}, \meas_{\cF}(\rhoh))
      + (1-p) \, \dist(\freq{\bz_{\ab}}, \meas_{\cFb}(\rhoh))
  \leq
    \eps \ .
  \]
  For any probability distribution $\Qh$ on $\cZ$ and any $a \subseteq
  \{1, \ldots, n\}$, let
  \[
    \cC(\Qh,a)
  :=
    \{\bz_{\ab}: \, (\bz, a) \in \cW
    \text{ and }
      \freq{\bz_{a}} = \Qh \} \ .
  \]
  We first show that
  \begin{equation} \label{eq:setsizeboundqm}
    \log(|\cC(\Qh,a)|)
  \leq
    |\ab| \, \fmax{H}(\cB(\Qh)) + \log(|\ab|) (|\cZ|-1) \ ,
  \end{equation}
  for any $\Qh$ and $a$. It follows from the definition of the set
  $\cC(\Qh,a)$ that for any $\bzp \in \cC(\Qh,a)$ there exists $\rhoh
  \in \cR$ such that $ \dist(\Qh, \meas_{\cF}(\rhoh)) \leq \eps / p $
  and $ \dist(\freq{\bzp}, \meas_{\cG}(\rhoh)) \leq \eps / (1-p) \ .
  $ which directly implies $\freq{\bzp} \in \cB(\Qh)$ and thus
  \[
    H(\freq{\bzp}) \leq r:= \fmax{H}(\cB)(\Qh)) \ .
  \]
  Hence, by Definition~\ref{def:typset}, $\bzp$ is contained in the
  $r$-typical set $\cT^{k}_{\cZ}(r)$ for $k:=|\ab|$. By
  Lemma~\ref{lem:typsetsize} the size of $\cT^{k}_{\cZ}(r)$ can not be
  larger than $2^{k r} |a|^{|\cZ|-1}$, from
  which~(\ref{eq:setsizeboundqm}) follows.

  Lemma~\ref{lem:quanttom} gives a lower bound for the probability
  that $(\bZ,A)$ is contained in $\cW$,
  \[
  \Prob[(\bZ, A) \in \cW] \geq 1-2^{|\cZ|+|\cZb|} e^{-\frac{n p^2}{8}} \ .
  \]
  Let the function $\mu$ be defined by
  \[
    \mu(\Qh,a)
  :=
    1-\Prob[\bZ_{\Ab} \in \cC(\freq{\bZ_A},A)|\freq{\bZ_A}=\Qh,A=a] \ .
  \]
  Then, since $\Prob[\bZ_{\Ab} \in \cC(\freq{\bZ_A},A)|] \geq
  \Prob[(\bZ, A) \in \cW]$, we obtain
  \[
    E[\mu(\freq{\bZ_A},A)]
  =
    1 - \Prob[\bZ_{\Ab} \in \cC(\freq{\bZ_A},A)|]
  \leq
    2^{2 q} e^{-\frac{n \eps^2}{8}} \ .
  \]
  On the other hand, from Lemma~\ref{lem:H0epscomp},
  \[
    H_0^{\mu(\Qh,a)}(\bZ_{\Ab}|\bZ_A=\Qh,A=a)
  \leq
    \log(|\cC(\Qh,a)|)
  \]
  for any $\Qh$ and $a$. Combining this with~(\ref{eq:setsizeboundqm})
  concludes the proof.
\end{proof}

\begin{corollary} \label{cor:quantentr}
  Let $\rho \in \cS(\cH^{\otimes n})$ be an $n$-partite state with
  density range $\cR \subseteq \cS(\cH)$, let $\cF = \{F_z\}_{z \in
    \cZ}$ be a POVM on $\cH$, let $\cFb$ be an orthogonal POVM on
  $\cH$, and let $A$ be a $p$-random selection on $\{1, \ldots, n\}$.
  Let $\bZ_A:=\ZVmeas_{\cF}^A(\rho)$ be the outcomes when measuring
  $\rho$ in $\cH_A$ with respect to $\cF$ and let $\rho_{\Ab}$ be the
  remaining quantum state in $\cH_{\Ab}$. Then, for any $\eps>0$,
  there exists a real valued function $\mu$ with
  \[
    E[\mu(\freq{\bZ_A},A)]
  \leq
    2^{\frac{\dim(\cH)+|\cZ|}{2}} e^{-\frac{n \eps^2}{16}}
  \]
  such that, for any probability distribution $\Qh$ on $\cZ$ and any $a
  \subseteq \{1, \ldots, n\}$,
  \[
    \qH_0^{\mu(\Qh,a)}(\rho_{\Ab}|\freq{\bZ_A}=\Qh, A=a)
  \leq
    |\ab| \, \fmax{H}(\cB(\Qh)) + \log(|\ab|) (\dim(\cH)-1) \ ,
  \]
  where $\cB(\Qh):= \cB_{\eps/(1-p)} (\meas_{\cFb}(\cR \cap
  \meas_{\cF}^{-1} (\cB_{\eps / p}(\Qh))))$.
\end{corollary}

\begin{proof}
  Since the POVM $\cFb = \{\Fb_z\}_{z \in \cZb}$ is orthogonal, we have
  $|\cZb| = \dim(\cH)$. According to Lemma~\ref{lem:quantentrbasic},
  there exists a function $\mub$ satisfying
  \[
    E[\mub(\freq{\bZ_A}, A)]
  \leq
    2^{\dim(\cH)+|\cZ|} e^{-\frac{n \eps^2}{8}}
  \]
  such that
  \begin{equation} \label{eq:Hzeroboundtemp}
    H_0^{\mub(\Qh, a)}(\freq{\bZ_{\Ab}}  |\freq{\bZ_A}=\Qh, A=a)
  \leq
    |\ab| \, \fmax{H}(\cB(\Qh)) + \log(|\ab|) (\dim(\cH)-1)
  \end{equation}
  holds. Let the function $\mu$ be defined by $\mu(\Qh,a):=\sqrt{2
    \mub(\Qh,a)}$. Using Jensen's inequality, we obtain
  \[
    E[\mu(\freq{\bZ_{A}},A)]
  \leq
    \sqrt{2 E[\mub(\freq{\bZ_{A}},A)]}
  \leq
    2^{\frac{\dim(\cH)+|\cZ|}{2}} e^{-\frac{n \eps^2}{16}} \ .
  \]
  On the other hand, since $\cFb$ is orthogonal,
  Lemma~\ref{lem:H0epsdec} implies that
  \[
    \qH_0^{\mu(\Qh,a)}(\rho_{\Ab}|\freq{\bZ_A}=\Qh, A=a)
  \leq
    H_0^{\mub(\Qh,a)}(\bZ_{\Ab}|\freq{\bZ_{A}}=\Qh, A=a)
  \]
  for any $\Qh$ and $a$ which, together
  with~(\ref{eq:Hzeroboundtemp}), concludes the proof.
\end{proof}

\begin{corollary}\label{cor:classtom}
  Let $\cZ$ be a set and let $A$ be a $p$-random selection on $\{1,
  \ldots, n\}$. Then, for any $n$-tuple $\bZ$ of random variables with
  range $\cZ$ and any $\eps>0$, there exists a real valued function
  $\mu$ with
  \[
    E[\mu(\freq{\bZ_{A}},A)]
  \leq
    2^{2 |\cZ|} e^{-\frac{n \eps^2}{2}}
  \]
  such that, for any probability distribution $\Qh$ on $\cZ$ and any
  $a \subseteq \{1, \ldots, n\}$,
  \[
    H_0^{\mu(\Qh, a)}(\bZ_{\Ab}|\freq{\bZ_{A}}=\Qh, A=a)
  \leq
    |\ab| \, \fmax{H}(\cB^{\eps / p (1-p)}(\Qh)) + \log(|\ab|) (|\cZ|-1) \ .
  \]
\end{corollary}

\begin{corollary} \label{cor:classtomcond}
  Let $\bX$ and $\bY$ be $n$-tuples of random variables with range
  $\cX$ and $\cY$, respectively, and let $A$ be a $p$-random selection
  on $\{1, \ldots, n\}$. Then, for any $\by \in \cY^n$ and $\eps>0$,
  there exists a real valued function $\mu$ with
  \[
    E[\mu(\freq{\bX_A|\bY_A},A)]
  \leq
    |\cY| 2^{2 |\cX|} e^{-\frac{n \eps^3}{2}}
  \]
  such that, for any channel $\Qh$ from $\cY$ to $\cX$ and for any set
  $a \subseteq \{1, \ldots, n\}$,
  \[
    H_0^{\mu(\Qh, a)}(\bX_{\Ab}|\freq{\bX_A|\bY_A} = \Qh, \bY=\by, A=a)
  \leq
    n \,(r + \eps |\cY| \log(|\cX|))
      + \log(n) (|\cX|-1) \ .
  \]
  where
  \[
    r:=\sum_{y \in \cY} \freq{\bY}(y) \fmax{H}(\cB^{{\eps / p (1-p)}}(\Qh(\cdot|y))) \ .
  \]
\end{corollary}

\begin{proof}
  Let $\cY'$ be the subset of $\cY$ containing all values $y$ such
  that $\freq{\by}(y) \geq \eps n$. For any $y \in \cY$, let
  \[
    a_y:=\{i: \by_i = y\} \ ,
  \]
  and, for any $y \in \cY'$, let $\mu_y$ be the function defined by
  Corollary~\ref{cor:classtom} applied to the tuple $\bX_{a_y}$. In
  particular, we have, for $y \in \cY'$, any probability distribution
  $\Qh'$ on $\cX$, and any $a \subseteq \{1, \ldots, n\}$,
  \[
    h_y
  :=
    H_0^{\mu_y(\Qh', a \cap a_y)}(\bX_{\Ab \cap a_y}|\freq{\bX_{A \cap a_y}}=\Qh', A=a)\leq
    |a_y| \, r_y + \log(n) (|\cX|-1) \ .
  \]
  where $r_y:=\fmax{H}(\cB^{\eps / p (1-p)}(\Qh))$.  On the other
  hand, for $y \in \cY - \cY'$, let
  \[
    h_y
  :=
    H_0(\bX_{\Ab \cap a_y}|\freq{\bX_{A \cap a_y}}=\Qh', A=a)
  \leq
    |a_y| \log(|\cX|)
  \leq
    n \eps \log(|\cX|)
  \]
  Applying Lemma~\ref{lem:entradd} yields
  \[
     H_0^{\mu(\Qh, a)}(\bX_{\Ab}|\freq{\bX_A|\bY_A} = \Qh, \bY=\by, A=a)
  \leq
    \sum_{y \in \cY} h_y
  \]
  for
  \[
    \mu(\Qh, a) := \sum_{y \in \cY'} \mu_y(\Qh(\cdot|y), a \cap a_y)
  \]
  from which the assertion follows.
\end{proof}

\subsection{Information Reconciliation}
\label{sec:IR}

\begin{lemma} \label{lem:infrec}
  Let $Z$ be a random variable with $H_{0}^{\eps}(Z) \leq r$ and let
  $F$ be a \tu{} hash function from $\cZ$ to $\{0,1\}^s$. Then there
  exists a guessing function $g$ such that
  \[
    \Prob[g(F, F(Z)) = Z] \geq 1-2^{-(s-r)} + \eps
  \]
\end{lemma}

For a proof, see e.g. \cite{RenWol04}.

\subsection{Privacy Amplification Against Quantum Adversaries}
\label{sec:PA}

We will use the following theorem proven in~\cite{KMR03}.

\begin{theorem} \label{thm:qmem}
  Let $Z$ be a random variable with $H_{\infty}^{\eps}(Z) \geq n$ and let
  $\rho \in \cS(\cH)$ be a density operator with $\qH_0^{\epsp}(\rho)
  \leq r$ which depends on $X$. Let $F$ be a \tu{} hash function from
  $\cZ$ to $\{0,1\}^s$ and let $W:=\ZVmeas_{\cG}(\rho)$ be the outcome
  of a measurement of $\rho$ with respect to an arbitrary POVM $\cG$
  which might depend on $F$. Then
  \[
    d(F(Z)|W F) \leq \frac{3}{4} 2^{-\frac{n-r-s}{2}} + \eps + \epsp \ .
  \]
\end{theorem}

\subsection{A Generic Quantum Key Distribution Protocol}
\label{sec:protocol}

In this section, we will describe the generic protocol and apply the results
from the previous sections to prove its security. To enhance the readability
of this exposition, we will restrict our attention to the asymptotic
behavior of the relevant quantities. The exact statements about eventual
constants may be taken directly from the lemmas that we refer to.

\label{subsec:generic:protocol} Let $\rho$ be a density operator on $(\cH_A
\otimes \cH_B)^{\otimes
  n}$. Let $\cF$ and $\cG$ be two POVMs on $\cH_A$ and let $\cF'$ and
$\cG'$ be two POVMs on $\cH_B$. Let $T$ and $T'$ be two $p$-random selections
on $\{1, \ldots, n\}$. For any $i \in \{1, \ldots, n\}$, let $X_i$ be the
outcome of a measurement of the subsystem $(\cH_A)_i$ with respect to $\cF$,
if $i \in T$, or with respect to $\cG$, otherwise. Similarly, let $Y_i$ be
the outcome of a measurement of $(\cH_B)_i$ with respect to $\cF'$, if $i \in
T'$, or with respect to $\cG'$, otherwise. Let $S$ be a $p$-random selection
on $\overline{T_A}$.

For the following asymptotic analysis, we assume that $p =
\Theta(n^{-\alpha})$ for some $\alpha \in (0, 1)$. In particular, $p n $
grows less than linearly in $n$.

\subsubsection{Parameter Estimation}

The goal of this protocol phase is to estimate the parameters used for
the subsequent information reconciliation and privacy amplification
phase. In particular, Alice and Bob have to determine the minimum
length $r$ of the error correcting information needed and the maximum
length $s$ of the final key such that it is guaranteed to be secure.

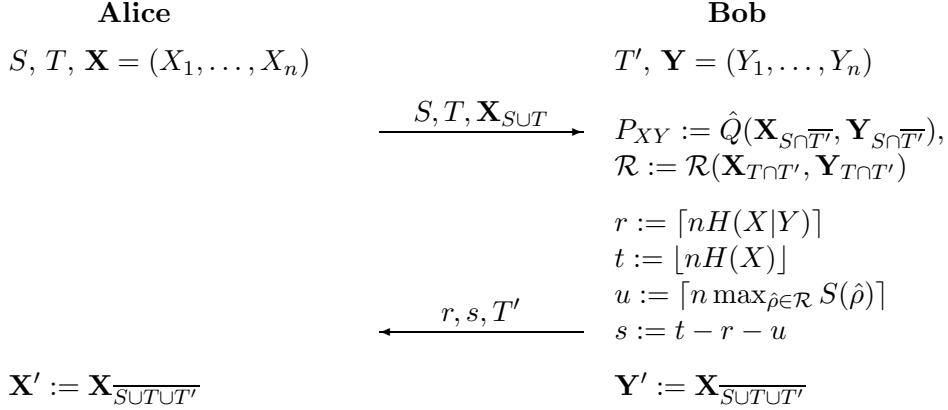
\begin{protocol}[0.95]{Alice}{Bob}
  \protno{$S$, $T$, $\bX=(X_1, \ldots, X_n)$}{$T'$, $\bY=(Y_1, \ldots, Y_n)$}
  \protright{}{$S, T, \bX_{S \cup T}$}
  {$P_{X Y}:=\Qh(\bX_{S \cap \overline{T'}}, \bY_{S \cap \overline{T'}})$,
   $\cR:=\cR(\bX_{T \cap T'}, \bY_{T \cap T'})$}
  \protleft{}{$r, s, T'$}
  {$r:=\lceil n H(X|Y) \rceil $\\
   $t:=\lfloor n H(X) \rfloor$\\
   $u:=\lceil n \max_{\rhoh \in \cR} S(\rhoh) \rceil$\\
   $s:=t-r-u$}
  \protno{$\bX':=\bX_{\overline{S \cup T \cup T'}}$}{$\bY':=\bX_{\overline{S \cup T \cup T'}}$}
\end{protocol}

The functions $\Qh$ and $\cR$ are defined as follows. Let $\bx = (x_1,
\ldots, x_k)$ and $\by = (y_1, \ldots, y_k)$ be two $k$-tuples.  Then
$\Qh(\bx, \by)$ is the frequency distribution $Q_{\bz}$ of the $k$-tuple
$\bz=((x_1, y_1), \ldots, (x_k, y_k))$. Similarly, $\cR:=\cR(\bx, \by)$ is
the set of density operators on $\cH_A \otimes \cH_B$ such that the outcomes
of a measurement of any $\rhoh \in \cR$ with respect to $\cF \otimes \cF'$
are distributed according to $\Qh(\bx, \by)$.

Note that $T \cap T'$ is a $p^2$-random selection on $\{1, \ldots, n\}$ and
that $S \cap \overline{T'}$ is a $p^2 (1-p)$-random selection on $\{1,
\ldots, n\}$. Corollary~\ref{cor:classtom} implies that
\begin{equation} \label{eq:Hzeroone}
  H^{\eps}_0(\bX'|\bY=\by, C)
\leq
  H^{\eps}_0(\bX_{\overline{S \cap \overline{T'}}})
  =
n H(X|Y) + o(n)
\end{equation}
holds for $\eps$ exponentially small in $n$. Similarly,
Lemma~\ref{lem:Hinfex} implies
\begin{equation} \label{eq:Hinfone}
  H^{\eps}_{\infty}(\bX'|C)
=
  H^{\eps}_{\infty}(\bX_{\overline{T \cap T'}}) + o(n)
=
  n H(X) + o(n) \ .
\end{equation}

\subsubsection{Information Reconciliation}

Let $n' := n - |S \cup T \cup T'|$ be the length of the tuples $\bX'$ and
$\bY'$, and let, for some $r' \leq n'$, $\cH_{r'}:=\cH(\cX^{n'} \to
\{0,1\}^{r'})$ be the set of \tu{} hash functions mapping $\bX'$ to $r'$
bits.

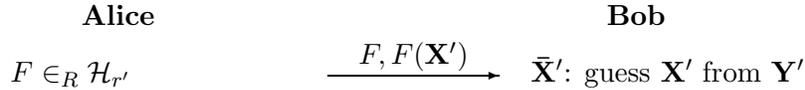
\begin{protocol}{Alice}{Bob}
  \protright{$F \in_R \cH_{r'}$}{$F, F(\bX')$}
  {$\bXb'$: guess $\bX'$ from $\bY'$}
\end{protocol}

It follows from Lemma~\ref{lem:infrec} and~(\ref{eq:Hzeroone}) that for some
$r' = r + o(n) \leq n H(X|Y) + o(n)$, $\bX' = \bXb'$ holds except with
probability exponentially small in $n$.  Moreover, since $F$ is independent
of $\bX'$ and since $H_0(F(X')) = r'$, Lemma~\ref{lem:Hinfsub} together
with~(\ref{eq:Hinfone}) implies
\begin{equation} \label{eq:Hinftwo}
  H_{\infty}^{\eps'}(\bX'|C,C')
\geq
  H^{\eps}_{\infty}(\bX'|C) - r' + o(n)
=
  n H(X) - r' + o(n)
\end{equation}
where $C':=(F,F(\bX'))$ are the messages sent by Alice during the information
reconciliation protocol.

\subsubsection{Privacy Amplification}
\label{subsubsection:PA} Let $\cP$ be the set of permutations of the $n'$
elements of $\bX'$ and for some $s' \leq n'$ let $\cH_{s'}:=\cH(\cX^{n'} \to
\{0,1\}^{s'})$ be a \tu{} hash function mapping $\bX'$ to $s'$ bits.

\begin{protocol}{Alice}{Bob}
  \protright{$P \in \cP, G \in_R \cH_{s'}$}{$P, G$}{}
  \protno{$S:=G(\bX')$}{$S':=G(P(\bXb'))$}
\end{protocol}

Since $\bX' = \bX''$ holds except with probability exponentially small in
$n$, we have $S = S'$.

It follows from Corollary~\ref{cor:quantentr} and~(\ref{eq:Hinftwo}) that for
some $s' = s + o(n)$,
\[
  H_{\infty}^{\eps'}(\bX'|C,C') - H_0^{\eps}(\rho) - s'
\geq
  n H(X) - r' - n \max_{\rhoh \in \cR} S(\rhoh) - s' + o(n)
\]
is smaller than zero.  Theorem~\ref{thm:qmem} thus implies that the knowledge
of Eve about the key $S$ is negligible.

Note that the length $s'$ of the final key is $t-r-u + o(n)$.  The
rate $R$ of this generic protocol is thus given by
\begin{equation} \label{eq:rate}
  R = I(X;Y) - \max_{\rhoh \in \cR} S(\rhoh) \ .
\end{equation}

Note further that we can carry out the same analysis for the difference
$H_{\infty}^{\eps'}(\bX'|C,W) - H_0^{\eps}(\rho|W)$ where we condition on
additional information $W$. This might improve the rate $R$ with a clever
choice of the information $W$ as we will see in the next section.

\section{Examples}
\label{sec:examples} In the following we will illustrate our result by
calculating the secret key rate and the tolerable error rates for common
quantum key distribution protocols.

\subsection{BB84 (The Four--State Protocol)}
The BB84 quantum key distribution protocol~\cite{BB84} belongs to the class
of so-called \emph{prepare and measure} protocols. In this protocol, Alice
chooses randomly, with probability $(1-p)$, the first out of a set of two
conjugate bases of a qubit, the second basis is chosen with probability $p$.
She then prepares one of the orthogonal basis states, each chosen with equal
probability, and sends the quantum state to Bob.~\footnote{The original
proposal by Bennett and Brassard fixes $p=\half$. A more efficient protocol,
which achieves twice the key generation rate of the original proposal, can be
obtained by choosing the two bases with different probabilities~\cite{LCA00}.
We choose $p=\Theta(n^{-\alpha})$, for $\alpha \in (0,1)$ and $n$ the number
of transmitted qubits (see also section \ref{subsec:generic:protocol}) . }

The BB84 protocol can be regarded as an entanglement based protocol and is in
this version known as BBM92~\cite{BBM92}. The preparation stage on Alice's
side is then given by a measurement on one half of an entangled quantum state
whose second part is sent off to Bob. The relevant quantum state $\rho$ is a
two qubit state, $\rho \in S(\mathcal{C}^2 \otimes \mathcal{C}^2)$ and we
denote measurement basis one by $\{\ket{0},\ket{1}\}$ and basis two by
$\{\ket{+},\ket{-}\}$, where $\ket{\pm}:=\frac{1}{\sqrt{2}}\left( \ket{0}\pm
\ket{1}\right)$. It is understood that e.g. $\bra{01} \rho \ket{01}$
corresponds to the probability that Alice obtains outcome $0$ and Bob outcome
$1$ when both choose to measure in the first basis. We further identify
$\ket{+}$ with outcome $0$ and $\ket{-}$ with outcome $1$.

After the phase, where the transmission of the quantum states and the
measurements have been finished, both parties publicly announce the bases in
which they conducted their measurements. They discard the cases in which they
did not measure in the same basis. On a small subset of the remaining data,
they compare a small part of the string to obtain an estimate of the error
rate. Let us assume that the error rate for measurements in both basis are
the same and equal to $\epsilon$. If this is not the case, Alice and Bob can
always randomly flip some of the bits of the set with the lower error rate in
order to make the error probabilities of both sets equal.

The entropy of Alice's string $X$ equals $H(X)=1$ and the conditional entropy
of $X$ given $Y$ is given by $H(X|Y)=h(\epsilon)$. The von Neumann entropy of
$\rho$ can be estimated as follows. Note that for all projective measurements
on $\rho$ with outcome given by a random variable $Z$, $H(Z)\geq S(\rho)$.
Using Alice's and Bob's data, we want to construct the data that a Bell
measurement, saved in the random variable $Z$ had resulted in. Let us define
the Bell states
\[ \ket{\psi^\pm}=\frac{1}{\sqrt{2}}\left(\ket{00}\pm \ket{11}\right) \quad
\text{and} \quad \ket{\phi^\pm}=\frac{1}{\sqrt{2}}\left(\ket{01} \pm
\ket{10}\right)\]

and express the probabilities of $Z$, denoted by $\lambda_i$, in terms of the
probabilities of measurements in basis one and basis two:

\begin{eqnarray}
\label{datastart}
\lambda_1&:=&\bra{\psi^+}\rho
\ket{\psi^+}=\bra{++}\rho\ket{++}+\bra{--}\rho\ket{--}-\bra{\phi^+}\rho\ket{\phi^+}\\
\lambda_2&:=&\bra{\psi^-}\rho
\ket{\psi^-}=\bra{+-}\rho\ket{+-}+\bra{-+}\rho\ket{-+}-\bra{\phi^-}\rho\ket{\phi^-}\\
\lambda_3&:=&\bra{\phi^+}\rho\ket{\phi^+}=\bra{01}\rho\ket{01}+\bra{10}\rho\ket{10}-\bra{\phi^-}\rho\ket{\phi^-}\\
\lambda_4&:=&\bra{\phi^-}\rho\ket{\phi^-} \ . \label{dataend}
\end{eqnarray}

The symmetric error probability $\epsilon$ yields

\begin{eqnarray*}
\bra{00}\rho\ket{00}+\bra{11}\rho\ket{11}&=&1-\epsilon\\
\bra{++}\rho\ket{++}+\bra{--}\rho\ket{--}&=&1-\epsilon\\
\bra{+-}\rho\ket{+-}+\bra{-+}\rho\ket{-+}&=&\epsilon\\
\bra{01}\rho\ket{01}+\bra{10}\rho\ket{10}&=&\epsilon
\end{eqnarray*}

and can be inserted into eqs. (\ref{datastart})-(\ref{dataend}). We obtain
\begin{eqnarray}
\label{eq-BB84-largestEV-start}
\lambda_3&=&\epsilon-\lambda_4 \\
\lambda_2&=&\epsilon-\lambda_4\\
\lambda_1&=&1-\epsilon-\lambda_3=1-2 \epsilon+\lambda_4 \ .
\label{eq-BB84-largestEV-end}
\end{eqnarray}

It remains to find the value of the free parameter $\lambda_4\in
[0,\epsilon]$ such that $H(Z)$ is maximized. It can easily be shown that this
is the case for $\lambda_4=\epsilon^2$ with $H(Z)=2 h(\epsilon)$.

The rate $R$ of the protocol according to eq. (\ref{eq:rate}) is given by

\begin{equation*}
R=H(X)-H(X|Y)-H(Z)=1-3h(\epsilon)
\end{equation*}

  The security threshold is the highest value of $\epsilon$ such that the rate $R$ is
positive and is henceforth the solution to the equation $1-3h(\epsilon)=0$.
We obtain $\epsilon \approx 0.061$ which corresponds to a $6.1\%$ bit error
rate. Conversely, there exists a quantum state $\rho$ for which this rate is
achieved and it is given by the mixture
\begin{eqnarray}
\label{Bell-diagonal} \rho=\lambda_1 \proj{\psi^+}+\lambda_2
\proj{\psi^-}+\lambda_3 \proj{\phi^+}+\lambda_4 \proj{\phi^-}
\end{eqnarray}
\par

Making use of the remark at the end of section \ref{subsubsection:PA}, we can
improve this security threshold. To do so, we introduce a random variable
$W=X \oplus Y$, which contains the information about the error positions. The
min entropy of the string $X$ does not decrease, whereas the size of the
quantum data does, thus improving the key rate $R$. This can be seen as
follows: given the fact that Alice and Bob measured in bases number one/two
and that an error/no error and has occurred, the quantum system can be
divided into 4 subsystems. The subsystems in the case of one error/no error
contain a fraction of $\frac{\epsilon}{2}$ and $\frac{1-\epsilon}{2}$ of the
total number of qubits, respectively. For each of the systems the entropy can
be estimated separately. If no error occurred we obtain $h(\frac{1-2
\epsilon+\lambda_4}{1-\epsilon})$ and if an error occurred we get
$h(\frac{\epsilon-\lambda_4}{\epsilon})$. Averaging over the four systems
gives

\begin{equation*} (1-\epsilon)h\left(\frac{1-2 \epsilon+\lambda_4}{1-\epsilon}\right)+\epsilon
h\left(\frac{\epsilon-\lambda_4}{\epsilon}\right)=H(Z)-h(\epsilon)
\end{equation*}

The key rate for BB84 is thus given by $R=1-2 h(\epsilon)$ and the security
threshold $\epsilon \approx 0.1100$ is the solution to the equation
$1-2h(\epsilon)=0$. The same rate has previously been obtained by Shor and
Preskill~\cite{SP00}.

\subsection{The Six--State Protocol}
The six--state protocol~\cite{Bru98, BPG99} is similar to the BB84 protocol,
but makes use of a third basis on either side. This additional basis is
defined as $\{\frac{1}{\sqrt{2}}(\ket{0}+i \ket{1}),
\frac{1}{\sqrt{2}}(\ket{0}-i \ket{1})\}$ and conjugate to the other
bases~\footnote{Note that we can choose bases two and three e.g. with
probability $\frac{p}{2}$ each}.  This protocol admits higher symmetry, since
the six states that are sent are symmetrically distributed on the Bloch
sphere. Similarly to the derivation of eqs.
(\ref{eq-BB84-largestEV-start})-(\ref{eq-BB84-largestEV-end}), we easily
derive the following additional constraint on the eigenvalues
\begin{equation*}
\lambda_3=\epsilon-\lambda_2
\end{equation*}
which results in $\lambda_1=1-3/2 \epsilon$ and $\lambda_i=\epsilon/2$ for $i
\in \{2, 3, 4\}$ corresponding to a security threshold of $6.8\%$ with
corresponding state

\begin{eqnarray}
\label{depolarised} \rho&=&\lambda_1 \proj{\psi^+}+\lambda_2
\proj{\psi^-}+\lambda_3 \proj{\phi^+}+\lambda_4 \proj{\phi^-}\\
&=&(1-2\epsilon)\proj{\psi^+}+ 2\epsilon \frac{\opone}{4}
\end{eqnarray}

Another way to derive this result uses the average fidelity $\bar{F}$ of a
qubit quantum channel. It can be shown~\cite{BOSBJ02} to be equal to the
average fidelity of the six states. Here, the fidelity of each state equals
$1-\epsilon$ and therefore $\bar{F}=1-\epsilon$. $\bar{F}$ and the
entanglement fidelity $F_e$ are related by the formula
$F_e=\frac{3\bar{F}-1}{2}$\cite{HHH99} which also leads to the $6.8\%$ by use
of the quantum Fano inequality.

The given bounds for BB84 and the six--state protocol on the maximal entropy
of $\rho$ are optimal, since Eve can simply prepare the state
eq.(\ref{Bell-diagonal}). Even if we consider Alice preparing the particles
and sending them off to Bob, we cannot achieve a better bound. This is
because the state eq. (\ref{Bell-diagonal}) can be effected by Eve with the
following strategy: apply the Pauli matrix $\sigma_z$ with probability
$\lambda_2$ ($\sigma_y$ with $\lambda_3$ and $\sigma_x$ with $\lambda_4$) on
the sent quantum state and with probability $\lambda_1$ take no action.

By conditioning on the random variable $W=X \oplus Y$, however, we can
improve the security threshold in a similar manner as we did in the BB84
analysis. This leads to a value of $\epsilon \approx 0.1262$ for the
six--state protocol, which coincides with the result of an earlier
calculation by Lo~\cite{Lo01} based on a result by Bennett {\it et
al.}~\cite{BDSW96}.

\subsection{B92}
In 1992, Bennett~\cite{Ben92} suggested a protocol for quantum key
distribution that belongs to the class of prepare and measure protocols
differs, however, significantly from BB84 and the six--state protocol. In the
specification of the protocol, known as B92, Alice sends one of two
non--orthogonal quantum states, which we will denote by $\ket{u_\pm}$, to
Bob. He chooses randomly to measure in one of two von Neumann measurements.
The first measurement consists of the vectors $\{\ket{u_-},
\ket{\tilde{u}_-}\}$, where $\ket{\tilde{u}_-}$ is orthogonal to $\ket{u_-}$.
Similarly, the second measurement is given by $\{\ket{u_+},
\ket{\tilde{u}_+}\}$ with $\ket{\tilde{u}_+}$ orthogonal to $\ket{u_+}$. Bob
announces acceptance if he obtains outcomes corresponding to
$\ket{\tilde{u}_\pm}$, otherwise both parties discard the values that they
recorded.

Alice records the bit value $0/1$ if she sends $\ket{u_+}/ \ket{u_-}$ and Bob
jots down the value $0/1$ if he obtains
$\ket{\tilde{u}_-}/\ket{\tilde{u}_+}$. We will assume throughout the analysis
that Alice sends each quantum state with equal probability and Bob chooses
randomly and with equal probability between his two measurements.

Note that in the case of perfect transmission, the strings, conditioned upon
acceptance are identical and randomly distributed. We will now proceed to
show how one can apply our generic security proof to this specific protocol
in the presence of noise. To do so, we need to estimate the expressions
in~(\ref{eq:rate}) where $\cR$ is the sets of possible quantum states
conditioned on the event that Alice and Bob accept.  As in the analysis of
BB84 and the six--state protocol, we will condition on an additional random
variable, which equals the XOR of Alice's and Bob's bits after acceptance.

For the following analysis, let $p_{x y}$ for $x, y \in \{0,1\}$, be the
probability that Alice and Bob accept a particle and that they have the bit
values $x$ and $y$, respectively. We can without loss of generality assume
that $p_{0 0} = p_{1 1}$ and $p_{0 1} = p_{1 0}$ (Alice and Bob can simply
abort the protocol if this is not the case).

Let $\{\ket{0}, \ket{1}\}$ be an orthonormal basis and write
\begin{eqnarray*}
\ket{u_\pm}=\beta \ket{0}\pm \alpha \ket{1}\\
\ket{\tilde{u}_\pm}=\alpha \ket{0}\mp \beta \ket{1}
\end{eqnarray*}
with $\alpha \in (0,\frac{1}{\sqrt{2}})$ and $\beta=\sqrt{1-\alpha^2}$. The
interaction of the transmitted quantum states with the environment or a
possible eavesdropper, Eve, is given by
\begin{eqnarray} \label{eq:evolution}
\ket{u_\pm}\ket{e} \mapsto \ket{\Psi_\pm}:=\sqrt{1-\delta}
\ket{u_\pm}\ket{e_\pm}+ \sqrt{\delta}\ket{\tilde{u}_\pm}\ket{\tilde{e}_\pm} \
,
\end{eqnarray}
where $\delta = 4 p_{0 1} = 4 p_{1 0}$. (Note that the factor $4$
results from the random choices of Alice and Bob.)

The evolution in equation eq. (\ref{eq:evolution}) is unitary which implies
the important constraint
\begin{equation*}
\braket{u_+}{u_-}=\braket{\Psi_+}{\Psi_-} \ .
\end{equation*}
This constraint reads in its expanded form
\begin{equation} \label{eq:unitarity}
\begin{split}
\beta^2-\alpha^2=&\ (1-\delta)(\beta^2-\alpha^2)\braket{e_+}{e_-}\\
        &+\sqrt{(1-\delta)\delta} \ 2 \alpha \beta
        \left(\braket{e_+}{\tilde{e}_-}+\braket{\tilde{e}_+}{e_-}\right)\\
        &+\delta (\alpha^2-\beta^2)\braket{\tilde{e}_+}{\tilde{e}_-} \ .
\end{split}
\end{equation}
Without loss of generality we can take $\braket{e_+}{e_-}$ to be real. Eve's
quantum states, given the outcome was accepted by Bob and that Alice and Bob
have the same bit value, are denoted by $\ket{f_\pm}$. In the case of an
error and acceptance, we write $\ket{\tilde{f}_\pm}$, where $\pm$ denotes
Alice's bit value $0/1$. One easily obtains
\begin{eqnarray} \label{eq:Eves-states}
\ket{f_\pm}&:=&\frac{\braket{\tilde{u}_\mp}{\Psi_\pm}}{\sqrt{\gamma}}
=\frac{\sqrt{1-\delta} \ 2 \alpha \beta
\ket{e_\pm}+\sqrt{\delta}(\alpha^2-\beta^2)\ket{\tilde{e}_\pm}}{\sqrt{\gamma}}\\
\ket{\tilde{f}_\pm}&:=&\frac{\braket{\tilde{u}_\pm}{\Psi_\pm}}{\sqrt{\delta}}=\ket{\tilde{e}_\pm}
\end{eqnarray}
where $\gamma = 4 p_{0 0} = 4 p_{1 1}$ is given by the probability that Alice
and Bob have a correct value,
\begin{eqnarray} \label{eq:accprob}
\gamma=(1-2 \delta)(2\alpha \beta)^2+2 \sqrt{(1-\delta)\delta} 2\alpha
\beta (\alpha^2-\beta^2) \re \braket{e_\pm}{\tilde{e}_\pm}+\delta \ .
\end{eqnarray}
Eve's density matrices, conditioned on the correctness/ falseness of the
accepted bit, are given by
\begin{eqnarray*}
\sigma&:=&\half \left( \proj{f_+}+\proj{f_-}\right) \\
\tilde{\sigma}&:=&\half \big( \proj{\tilde{f}_+}+\proj{\tilde{f}_-}
\big)=\half \big( \proj{\tilde{e}_+}+\proj{\tilde{e}_-} \big)
\end{eqnarray*}
and have eigenvalues $\frac{1\pm |\braket{f_+}{f_-}|}{2}$ and $\frac{1\pm
|\braket{\tilde{e}_+}{\tilde{e}_-}|}{2}$, respectively. Every estimate for
the scalar products $\braket{f_+}{f_-}$ and
$\braket{\tilde{e}_+}{\tilde{e}_-}$ thus leads to an estimate of the entropy
of $\sigma$ and $\tilde{\sigma}$. $\braket{f_+}{f_-}$ takes the form
\begin{eqnarray*}
\braket{f_+}{f_-}&=&\frac{
(1-\delta)\braket{e_+}{e_-}-(\beta^2-\alpha^2)^2}{\gamma}
\end{eqnarray*}
where we made use of eq.~(\ref{eq:unitarity}) to simplify the expression in
the nominator.

It thus remains to find an estimate for $\braket{e_+}{e_-}$. This will be
done by use of the unitarity constraint, eq.~(\ref{eq:unitarity}). In
particular, we can choose $\delta$ such that $\braket{e_+}{e_-}$ is
sufficiently close to one. For small $\re \braket{e_\pm}{\tilde{e}_\pm}$, we
thus derived lower bound on the scalar product $\braket{f_+}{f_-}$. Note that
$\delta$ and $\gamma$ can be derived from the probabilities $p_{x y}$ which
are determined by Alice and Bob. Together with eq.~(\ref{eq:accprob}), this
gives an estimate for $\re \braket{e_\pm}{\tilde{e}_\pm}$. Using the trivial
bound $S(\tilde{\sigma}) \leq 1$, this suffices to find a bound for the rate
of B92 according to eq.~(\ref{eq:rate}).

As a specific example let us consider the depolarizing channel
\begin{equation*}
\rho \rightarrow (1-p) \rho+\frac{p}{3}\sum_i \sigma_i \rho \sigma_i \ .
\end{equation*}
It is easy to compute the quantities $p_{x y}$ for this channel. In
particular, we obtain
\[
  p_{0 1} = p_{1 0} = p/6
\]
and
\[
  p_{0 0} = p_{1 1} = \frac{1}{4} (1-\frac{4}{3} p) (2 \alpha \beta)^2 + \frac{2}{3} p
\]
i.e., $\delta=\frac{2}{3} p$. Using eq.~(\ref{eq:accprob}) and
$\gamma=4p_{00}=4 p_{11}$, we have $\re \braket{e_\pm}{\tilde{e}_\pm}=0$. The
error rate conditioned on acceptance, is thus given by
\begin{equation*}
\epsilon=\frac{\delta}{(1-2 \delta)\eta+2 \delta} \qquad \mbox{with }
\eta:=(2\alpha \beta)^2 \ .
\end{equation*}
From $\re \braket{e_+}{\tilde{e}_+}=0$ follows $\re
\braket{e_+}{\tilde{e}_-}\leq \sqrt{1-|\braket{e_+}{e_-}|^2}$ which we insert
into eq. (\ref{eq:unitarity}). We therefore have an estimate of the terms
proportional to $\sqrt{\delta}$. For the third term of the right hand side of
eq. (\ref{eq:unitarity}), we use take the trivial estimate $\re
\braket{\tilde{e}_+}{\tilde{e}_-}\geq -1$. Altogether we have
\begin{equation*}
\nu \ (1-\braket{e_+}{e_-})^2\leq 1-\braket{e_+}{e_-}^2 \quad \mbox{with }
\nu:=\frac{(1-\delta)(1-\eta)}{4\delta \eta} \ .
\end{equation*}
The valid solutions of this quadratic expression are given by
\begin{equation*}
\braket{e_+}{e_-} \geq \frac{\nu-1}{\nu+1}
\end{equation*}
and directly lead to an estimate for $S(\sigma)$. Using
$S(\tilde{\sigma})\leq 1$ we obtain an estimate for the entropy of the
quantum state of Alice and Bob conditioned on the random variable $W$.
The total rate is given by
\begin{eqnarray*}
R=\frac{ (1-2 \delta)\eta +2\delta}{2} \big( 1-h(\epsilon)-\epsilon
-(1-\epsilon) h(x) \big)
\end{eqnarray*}
where
\begin{eqnarray*}
 x:=\frac{(1-5\delta)(1-\delta)\eta (1-\eta)}{(\delta+(1-2
\delta)\eta)((1-\delta)-(1-5\delta)\eta)}
\end{eqnarray*}

The highest security threshold $p$ is obtained for $\alpha \approx 0.38$ and
equals $p \approx 0.036$. This is a slight improvement of the previously
obtained security threshold $p \approx 0.034$ by Tamaki, Koashi and
Imoto~\cite{TKI03}.

\section{Conclusion}
In this paper we have presented a security proof for a generic quantum key
distribution protocol. The protocol requires only single particle
measurements on Alice's and Bob's sides and uses one-way information
reconciliation and privacy amplification to extract a secret key from the raw
data. In our proof we estimate the amount of classical correlation contained
in Alice's and Bob's data and derive a bound on the quantum information,
which a possible adversary might have about this data. Subsequently, we apply
a recent result by K\"onig, Maurer and Renner~\cite{KMR03} to ensure the
security of the privacy amplification stage.

Special cases of our protocol include entanglement based quantum key
distribution, such as E91, and prepare and measure schemes, such as BB84 or
the six state protocol. We were able to derive security thresholds of
$11.0\%$ bit error rate for BB84 (four--state protocol) and $12.6\%$ for the
six--state protocol, previously obtained by Shor and Preskill, and Lo,
respectively. Furthermore we have shown how our technique can be applied to
prove the security of B92. In the case of the depolarizing channel this leads
to a slight improvement of the security threshold that has been recently
obtained by Tamaki, Koashi and Imoto.

\section{Acknowledgements}
This work was supported in part by a grant from the Cambridge-MIT
Institute, A$^*$Star Grant No.\ 012-104-0040 and the EU under project
RESQ (IST-2001-37559). MC was supported by a DAAD
Doktorandenstipendium. RR was partially supported by the Swiss
National Science Foundation, project No.~20-66716.01.

\bibliographystyle{plain}

\end{document}